\documentclass[12pt]{article}
\usepackage[margin=1in]{geometry}
\usepackage{setspace}
\usepackage{amsmath,amssymb,bm}
\usepackage{multirow,color}
\usepackage{graphicx}
\usepackage{natbib}
\usepackage{chngcntr}
\def\bSig\mathbf{\Sigma}
\newcommand{\bd}{\bm{d}}
\newcommand{\bD}{\bm{D}}
\newcommand{\bt}{\bm{t}}
\newcommand{\bT}{\bm{T}}
\newcommand{\bX}{\bm{X}}
\newcommand{\bK}{\bm{K}}
\newcommand{\bN}{\bm{N}}
\newcommand{\bx}{\bm{x}}

\newcommand{\by}{\bm{y}}
\newcommand{\bZ}{\bm{Z}}
\newcommand{\bG}{\bm{G}}

\newcommand{\bk}{\bm{k}}

\newcommand{\bbeta}{\boldsymbol{\beta}}

\newcommand{\btheta}{\boldsymbol{\theta}}
\newcommand{\bphi}{\boldsymbol{\phi}}

\begin{document}

\title{\large \textbf{Child Mortality Estimation Incorporating Summary Birth History Data}}

\author{\normalsize Katie Wilson$^{1}$ and
Jon Wakefield$^{1,2}$ \\
\normalsize $^1$Department of Biostatistics, $^2$Department of Statistics, \\
University of Washington, Seattle, Washington\\
}

\date{}
\maketitle

\begin{abstract}
The United Nations' Sustainable Development Goal 3.2 aims to reduce under-5 child mortality to 25 deaths per 1,000 live births by 2030. Child mortality tends to be concentrated in developing regions where much of the information needed to assess achievement of this goal comes from surveys and censuses. In both, women are asked about their birth histories, but with varying degrees of detail.
Full birth history (FBH) data contain the reported dates of births and deaths of every surveyed mother's children. In contrast, summary birth history (SBH) data contain only the total number of children born and total number of children who died for each mother. Specialized methods are needed to accommodate this type of data into analyses of child mortality trends. We develop a data augmentation scheme within a Bayesian framework where for SBH data, birth and death dates are introduced as auxiliary variables. Since we specify a full probability model for the data, many of the well-known biases that exist in this data can be accommodated, along with space-time smoothing on the underlying mortality rates. We illustrate our approach in a simulation, showing that uncertainty is reduced when incorporating SBH data over simply analyzing all available FBH data. We also apply our approach to data in the Central region of Malawi.
We compare with the well-known Brass method.\\

\end{abstract}

%



%

\section{Introduction}
\label{s:intro}

An estimate of the under-five mortality rate (U5MR) over time is crucial for determining effectiveness of public health interventions and for better allocating resources. In countries without vital registration systems that track births and deaths, the data often come from surveys that are typically of one of two types: full birth history (FBH) and summary birth history (SBH). In surveys that collect FBH data, women are asked to recall the birth and, if applicable, death dates for each of their children. In contrast, in surveys that collect SBH data, women are only asked for the total number of children they have had and the number of those children that died. Given the temporal information contained in FBH data, it is relatively straightforward to obtain estimates of child mortality and U5MR using survey or model-based approaches 
\citep{mercer2015space,pezzulo:etal:16,rutstein:06,wakefield:etal:18}. 
Unlike FBH data, SBH data 
do not directly provide temporal information on when births and deaths occurred, and thus require more specialized methods. 
Overall, SBH data are easier and faster to collect, and thus widely available. In a recent study of U5MR in Africa, approximately 44\% of the surveys contained only SBH data and 90\% of the births were from SBH 
\citep{golding:etal:17}.

\subsection{Existing SBH Methods}

A common approach to obtaining U5MR estimates based on SBH data is to use variations of the Brass method 
\citep{brass:64,feeney:80,trussell:75}, which involve model life tables. This approach provides an estimate of U5MR for five-year age groups of women and additionally assigns each estimate a reference time in the past using a series of complex formulas. These formulas make use of quantities that are observed in SBH data: the number of children that died, the number of children born, and number of women surveyed along with women's age. 
Since this approach does not use a statistical model, it is not immediately clear how to obtain estimates of uncertainty. One proposed approach is to use the jackknife 
\citep{pedersen:2012}. 
A major shortcoming of the Brass method is that the mortality rates in the most recent time periods are based on the youngest women. Children born to younger mothers tend to have worse outcomes, resulting in bias. Therefore, these estimates are typically excluded from analyses. 
Alternatively, Brass-type methods using time since first birth or marriage are used instead of age of mother
\citep{hill:99,manualX:83}.
Currently, the United Nations produce national estimates of U5MR that incorporate SBH data via indirect estimates obtained from the time since first birth variant of the Brass method 
\citep{alkema:2014a,hill:12}. 
\cite{hill:12} summarizes the historical approach of combining the estimates using a loess smoother in time. The bootstrap can be used to obtain uncertainty intervals \citep{alkema2012progress}. A more recent approach, proposed by 
\cite{alkema:2014a} and \cite{alkema:2014b}, combines the different sources of data and uses a Bayesian penalized regression spline to model trends over time. Uncertainty in the SBH data is incorporated by using the jackknife, or survey weights if microlevel data is available, and otherwise fix the uncertainty to a predetermined number.

Other methods 
\citep{hill:15,rajaratnam:2010} that do not rely on the demographic models used in the Brass method have been proposed, but critically do not assume a full probability model. 
\cite{rajaratnam:2010} describe a number of methods, including one that uses FBH data to derive empirical distributions of births and deaths prior to the survey and then matches SBH women to the relevant empirical distribution. This provides a yearly estimate of the ratio of children that died to children ever born. The ratio is then related via a logistic regression model to the probability that a child dies within five years, calculated using FBH data. In practice, FBH data from surveys in different countries and time periods are pooled together to build the regression model and empirical distributions. 
\cite{hill:15} propose a birth history imputation approach in which SBH women are matched to FBH women, who are typically available from an earlier survey in that country. Women are matched by age, number of births, and number of deaths. The FBH data 
are then used to impute births and deaths to the SBH women. 
This approach gave disappointing results when validating mortality estimates obtained from this method and comparing to estimates computed from a later FBH survey 
\citep{brady:2017,hill:15}. The authors attribute this to incompatibility of the SBH and FBH data, stemming from data quality issues with SBH data.

\subsection{Malawi Data Available}

As our motivating example, we consider retrospective birth history data taken between 2004 and 2015 in Malawi, from five surveys and a census. Specifically, we use FBH data from the 2004 Malawi Demographic and Health Survey (DHS), the 2010 Malawi DHS, the 2015 Malawi DHS, the 2006 Multiple Indicator Cluster Survey (MICS), and the 2013 MICS. We include surveyed women who are aged 15--49 at time of the survey. We use SBH data from the 2008 Malawi Census. From the census, we include women who are aged 25--49 at time of the survey. Excluding women under 25 who provide SBH data is consistent with other analyses that incorporate SBH data 
\citep{hill:12}.
The DHS and MICS use 2-stage stratified cluster sampling designs. The 2004 DHS was stratified by urban-rural classification with certain districts being oversampled. The 2006 MICS was stratified by district. The 2010 DHS, 2013 MICS, and 2015 DHS were stratified by urban-rural classification crossed with district.
We use available microdata, which is a 10\% sample of the 2008 Census.

We focus on the 9 districts in the Central region: Dedza, Dowa, Kasungu, Lilongwe, Mchinji, Nkhota Kota, Ntcheu, Ntchisi, and Salima. In this region, birth history information is available from 109,713 women (43\% of women had SBH information only). 
The number of clusters, women surveyed, births, and deaths for each of the surveys and census is in Table \ref{Tab:data_descrip}. 
Across all surveys and the census, the age of the mother at time of survey, district, and strata are available. The 5 surveys that contain FBH information also provide reported birth and death dates for each child. The census contains only the reported number of births and deaths for each woman. 
An exploratory analysis, comparing the age of surveyed women, total number of reported births, and ratio of number of deaths to number of births, is shown in Figure \ref{Fig:eda1}.

\begin{figure}[t]
  \centerline{\includegraphics[width=6.5in]{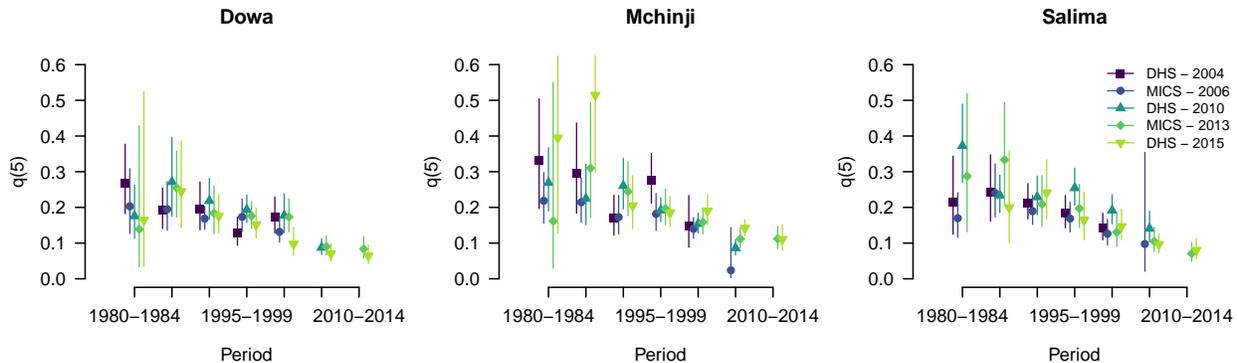}}
  \caption{HIV-adjusted weighted estimates of under-five mortality ($q(5)$) over time for 3 districts in central Malawi, by survey.  A color version of this figure can be found in the electronic version of the article.}
\vspace*{-3pt}
\label{Fig:de}
\end{figure}

Weighted estimates 
\citep{horvitz:thompson:52} and 95\% confidence intervals (CIs) going back to 1980 were computed for each of the FBH surveys and are shown in Figure \ref{Fig:de} (we include 3 districts for illustration, with the remaining 6 districts being given in Figure \ref{Fig:de}). Overall, under-five mortality is decreasing, but there is significant heterogeneity over districts and time periods. 

The organization of this paper is as follows. In Section \ref{Sec:method} we describe our method, which is based on data augmentation (DA). In  Section \ref{Sec:combomethod} we describe an alternative approach based on combining estimates from using weighted, also known as direct, estimators on FBH data, and indirect estimates from using the Brass method on SBH data. In
Section 
\ref{Sec:simulation} we show the usefulness of the approaches in a simulation. We apply our approach to the survey and census data from Malawi in Section \ref{Sec:results}. Finally, we conclude with a discussion in Section \ref{Sec:conclusion}.

\section{Data Augmentation Method \label{Sec:method}}
We propose using a DA approach within a Bayesian framework. This is implemented via a Markov chain Monte Carlo (MCMC) algorithm, where each iteration is divided into two major steps. In the first step, the missing birth and death dates (available for FBH data) are introduced as auxiliary variables for the SBH data. In the second step, mortality and fertility parameters are then updated conditional on this imputed FBH data and combined with the existing FBH data.

In modeling SBH data, we will model fertility rates and mortality rates, which we specify as probabilities. The forms for these models are heavily driven by context. We define the fertility rate as the probability a woman gives birth at age $m$ and time $t$ and denote it as $f(m,\bx(t))$, where $\bx(t)$ contains the covariates at time $t$ associated with fertility. The mortality rate, or probability a child dies between age $a$ and $a+1$, is denoted by $_1 q_a(\bx(t)) = q_a(1,\bx(t))$, where $\bx(t)$ again contains covariates at time $t$ associated with mortality. 


\subsection{Simple Example} 
\begin{figure}[t]
  \centerline{\includegraphics[width=6.5in]{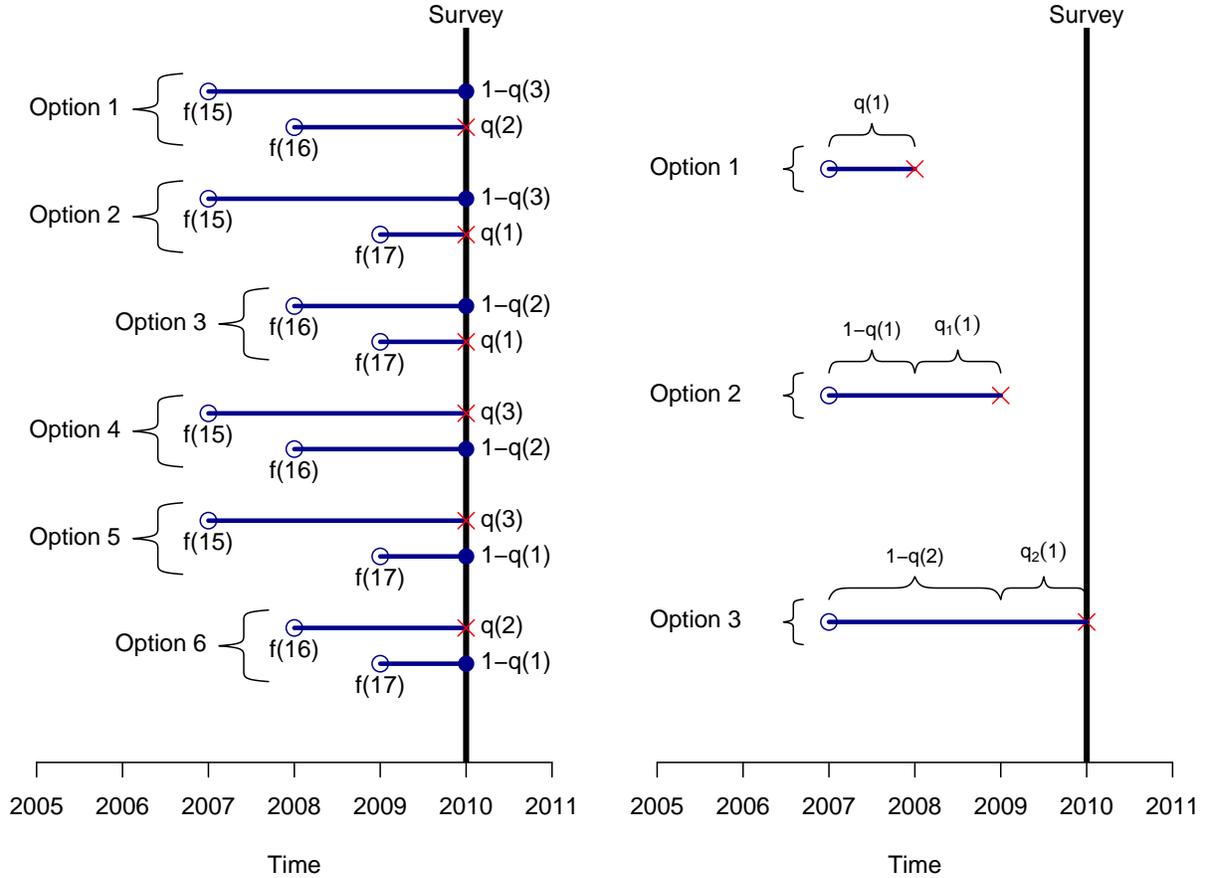}}
  \caption{Left: Possible birth years and death indicators for 2 children with mother's age $m_{surv}=18$, assuming 1 death. Fertility $f(m)=0$ for $m < 15$. Right: Possible death ages for a child born to a mother of age $m_b=15$. The {\color{red}$\times$}'s represent a death. Here, $q(a)$ is the probability of death by age $a$.  A color version of this figure can be found in the electronic version of the article.}
\vspace*{-3pt}
\label{Fig:method}
\end{figure}

To motivate our approach, we first consider a simple scenario where we suppose that fertility and mortality do not change with time. Suppose we survey a woman who is age 18 at the time of the survey and has had two children, one of whom has died. Further suppose that the youngest age she could have given birth was 15 and that it is not possible to have multiple births. Additionally, we will discretize time and work on a yearly scale. 
Therefore, assuming the woman could have given birth when she was 15, 16, or 17, there are 3 options for when the children were born relative to the woman's age (i.e., at ages 15 and 16, 15 and 17, or 16 and 17). We also know that one of the children has died. This means there are 6 options for when the children were born and which of the children died (e.g., the child born when the mother was 15 died and the child born at 16 survived). The probabilities of each of these 6 options will depend on the fertility rates and the mortality rates. A visual depiction of these scenarios is given in Figure~\ref{Fig:method}.

Now suppose we knew when the children were born and which of those children died (i.e., we selected one of the 6 scenarios). For the child who died, we can enumerate the options for age of death. In this case, suppose the woman was 15 when the child who died was born. Again considering a yearly time scale, 
there are 3 scenarios (i.e., the child died between ages 0 and 1, ages 1 and 2, or ages 2 and 3). The probability associated with each of these options is based on the mortality rates (see Figure \ref{Fig:method}). 

We have given an overview of the DA that links SBH information to FBH information for a simple example. 
This forms the backbone of the DA step of our method, and we now give details of the probability model that is used in this step.

\subsection{Step 1: Data Augmentation} 
Let $t$ denote calendar time in years. Consider a woman with $B$ births, $D$ deaths, and covariates $\bx(t)$. Let $m_{surv}$ be the age of the woman at time of survey and $t_{surv}$ be the year the survey was completed. Define $\bt_b = \{t_{b_1},\dots,t_{b_B}\}$ to be a vector of length $B$ which contains the (unknown) years of birth of the $B$ children. Define $\bd=\{d_1,\dots,d_B\}$ to be a vector of length $B$ which contains the ``death indicators'' for the children (i.e., $d_i=1$ if child $i$ dies). Therefore, $\sum_{i=1}^B d_i =D$.  It will also be helpful to define $m_{b_i}$ to be the age of the woman when she gave birth to child $i$. Note that $m_{b_i} = m_{surv} - (t_{surv} - t_{b_i})$.

The probability of children being born in particular years and surviving or dying by the time of the survey is proportional to the probability of giving birth to the children in those years, observing them to survive through the survey or die at some point prior to the survey, and not having children born in the other years:
\begin{align}
& \Pr\big( \bT_b = \bt_b, \ \bD = \bd
\big) \propto \Bigg(\prod_{i=1}^{B}{f(m_{b_i},\bx(t_{b_i}))} \nonumber \times \left[1- \prod_{a=1}^{t_{surv}-t_{b_i}}\left\{1-\! _{1} q_{a-1}(\bx(t_{b_i} + a - 1))\right\}\right]^{d_i} \nonumber \\
& \quad \times \left[\prod_{a=1}^{t_{surv}-t_{b_i}}\left\{1-\! _1q_{a-1}(\bx(t_{b_i} + a - 1))\right\}\right]^{1-d_i} \Bigg) \times \prod_{\substack{t \not \in \{t_{b_1},\dots,t_{b_B}\},\\ t < t_{surv}}}\{1-f(m,\bx(t))\}.
\label{eq:substep1}
\end{align}
This probability is conditional on $B$ births, $D$ deaths, covariates $\bx(t)$, survey taken in year $t_{surv}$, a woman of age $m_{surv}$ at time of survey, and functions $f(\cdot)$ and $q(\cdot)$, but for notational ease we do not explicitly state this.
As expressed, this probability does not allow for multiple births and assumes that births in separate years are independent. However, these assumptions could be relaxed.  

Given birth in year $t_{b_i}$, $d_i=1$, covariates $\bx(t)$, and the mortality function $q(\cdot)$, the probability that child $i$ dies in a particular year $t_{d_i}$ is proportional to the probability of surviving to year $t_{d_i}-1$ and subsequently dying in the next year,
\begin{align}
\Pr(T_{d_i} = t_{d_i}) \propto\!\,  _1q_{t_{d_i}-t_{b_i}-1}(\bx(t_{d_i}-1)) \prod_{a=1}^{t_{d_i} - t_{b_i} - 1 }\{1-\! _1 q_{a-1} (\bx(t_{b_i}+a-1))\}. \label{eq:substep2}
\end{align}

These are the necessary probabilities for updating the birth and death years for all SBH women's children (an example of which is shown in Figure \ref{Fig:method}). In terms of computation, it is possible to enumerate all options thus computing the denominators for  (\ref{eq:substep1}) and (\ref{eq:substep2}), and sample from the full posterior conditional distribution. This can be efficiently achieved by grouping women together that have the same age, number of births, number of deaths, and covariates so that the desired probabilities need only be computed once. However, computational gains can be made using a Metropolis-Hastings algorithm. This is especially useful for older women, who tend to have more children. One simple approach is to iterate through all of a woman's children and sample the birth years one-by-one, where $q(\cdot)$ on the left hand side represents the proposal probability,
\begin{align*}
q(T_{b_i} = t_{b_i}, d_i=1\ |\ i \leq D) & \propto f(m_{b_i},\bx(t_{b_i})) \times \left[1- \prod_{a=1}^{t_{surv}-t_{b_i}}\left\{1-\! _{1} q_{a-1}(\bx(t_{b_i} + a -1))\right\}\right],\\
q(T_{b_i} = t_{b_i}, d_i =0 \ |\ i > D) & \propto f(m_{b_i},\bx(t_{b_i})) \times \left[\prod_{a=1}^{t_{surv}-t_{b_i}}\left\{1-\! _1q_{a-1}(\bx(t_{b_i} + a -1))\right\}\right].
\end{align*}
In the data analysis, this independence chain proposal resulted in acceptance ratios $> 0.6$.

\subsection{Step 2: Parameter Updates}

After performing the DA step, the imputed FBH from the SBH is combined with available FBH and the fertility and mortality probabilities are updated. 
Let $\mbox{Y}(m,t)$ be an indicator for birth at woman's age $m$ and year $t$, then $\mbox{Y}(m,t)\ |\ f(m,\bx(t)) \sim \text{Bernoulli}\left(f(m,\bx(t))\right)$.

For the mortality model, we will use a discrete hazards model. Let $\mbox{Z}_a(t)$ be an indicator for death between age $a$ and $a+1$ in year $t$. The likelihood is $\mbox{Z}_a(t)\ |\ _1q_a(\bx(t)) \sim \mbox{Bernoulli}\left( _1q_a(\bx(t))\right)
$.
The U5MR for time $t$ and with covariates $\bx(t)$ is,
\begin{align*}
_5 q_0 (\bx(t)) = 1 - \prod_{a=0}^4\left\{1-\! _1q_a(\bx(t))\right\}.
\end{align*}
Forms for $f(\cdot)$ and $_1q_a(\cdot)$ are given in Sections 
\ref{Sec:simulation} and \ref{Sec:results} for the simulation and data analysis, respectively.

\section{Weighted Estimation with the Brass Method \label{Sec:combomethod}}

First, we briefly provide a review of the Brass method as it pertains to obtaining estimates of the U5MR and a description of how to obtain estimates of uncertainty. Lastly, we describe how to combine Brass and weighted estimates and incorporate an adjustment for HIV.

\subsection{The Brass Method}

Define $D_{m_{surv}}$, $B_{m_{surv}}$ and $d_{m_{surv}}=D_{m_{surv}}/B_{m_{surv}}$ to be, respectively, the total number of children dead, the total number of children born and  the proportion that died, to women aged ${m_{surv}}$ at the time of survey. Then
 \begin{align}
 E(d_{m_{surv}}|B_{m_{surv}}) = \int_{0}^{m_{surv}} c_{m_{surv}}(a) q(a)\text{d}a \label{eq:brassE}
 \end{align}
 where $c_{m_{surv}}(a)$ is the proportion of births to women who are $m_{surv}$ at the time of the survey $a$ years prior to the survey and $q(a)$ is the probability that a child born $a$ years before the survey dies before the survey. The Brass method treats (\ref{eq:brassE}) as deterministic, replacing the expectation on the left side with the observed proportion, $d_{m_{surv}}$.
 By the mean value theorem, there exists an $a^* \in (0, m_{surv})$ such that,
 $$d_{m_{surv}} = q(a^*) \int_0^{m_{surv}} c_{m_{surv}}(a)\text{d}a = q(a^*).$$
 The key idea of the Brass method is to identify $a^*$ and thus use the observed proportion of children dead to find $q(a^*)$ (Chapter 11 of Preston et al., 2000)\nocite{preston2000demography}.
 
\cite{brass:64} achieved this by using model life tables and a polynomial fertility model to numerically integrate (\ref{eq:brassE}) using 5-year age groups of women, $i= 15-19, 20-24,\dots, 45-49$. The proportion dead in age group $i$, denoted $\tilde{d}_i$, is then compared to a $\tilde{q}(a)$ curve obtained from model life tables to find the $a_i^*$ such that $\tilde{d}_i = \tilde{q}(a_i^*)$. These times, $a_i^*$, were then adjusted to whole number of years. For example, $i= 15-19,\ 20-24,\ 25-29,\ 30-34$ correspond approximately to $q(1),\ q(2),\ q(3),\ q(5)$ (see Table 4 of Brass, 1964). These correspondences are not exact; therefore, adjustment terms are needed. Ideally, $c_{i}(a)$ (the proportion of births to women in age group $i$) would be used. However, with those proportions unknown, fertility information from cohorts of women are used instead. The Brass method makes use of observed parity measures, taken to be the mean number of children born to women in each age group. Comparing parity measures across age groups can provide a sense of the earliness of fertility. Define $P_1$, $P_2$, and $P_3$ to be the mean number of children ever born to all women in age groups $15-19$, $20-24$, and $25-29$, respectively.  
\cite{trussell:75} developed the model   that is commonly used to adjust the observed proportion of children dead $d_i$ to $q(x)$,
\begin{align*}
 q(x) & = d_i \left(a_{1i} + a_{2i} \frac{P_1}{P_2} + a_{3i} \frac{P_2}{P_3}  \right),
 \end{align*}
 where the  set of coefficients, $a_{1i},\, a_{2i},$ and $a_{3i}$ were estimated via simulation. In fact, 4 sets of coefficients were derived, one for each of the 
 \cite{coale:demeny:83} regional model life tables: ``North'', ``West'', ``South'', and ``East''.
 
To acknowledge changing mortality, 
\cite{coale:trussell:77} 
assume mortality declines linearly. Using (\ref{eq:brassE}), the reference time $a^*$ would equal the mean length of time since birth of children. Again using simulation, 
\cite{coale:trussell:77} develop another formula involving parity measures and coefficients, $b_{1i},\, b_{2i},$ and $b_{3i}$, to identify a time for which the Brass estimate is most relevant,
 \begin{align*}
  t(x) & = b_{1i} + b_{2i} \frac{P_1}{P_2} + b_{3i} \frac{P_2}{P_3}.
  \end{align*}
  
Finally, to obtain estimates of $q(5)$, the Brass estimates $q(x)$ are converted using numbers based on life tables. To determine which model life table to follow (and thus which set of coefficients to use), we follow the suggestion of \cite{hill:2013} and first obtain direct estimates of $_1q_0$ and $_4q_1$ from the 2010 DHS and 2006 MICS (FBH surveys taken near to the time of the census). We compare the direct estimates at the regional level with the 4 Coale--Demeny regional model life tables and select the model life table that the direct estimates most closely follow (see Figure \ref{Fig:brass-models}); we use the ``North'' regional model life table.  The coefficients for both formulas and adjustment procedure can be found in the United Nations' \textit{Manual X} 
\citep{manualX:83}.

\subsection{Estimates of Uncertainty for the Brass Method}

Define $\hat{\theta}(t)^{\text{B},\star} = \text{logit}\left( _5\hat{q}_{0}(t)^\star\right)$ where $_5\hat{q}_{0}(t)^\star$ are the Brass (indirect) estimates. We assume,
\begin{align*}
\hat{\theta}(t)^{\text{B},\star} \sim N\left(\theta(p)^\star, V(t)^{\text{B},\star}\right)
\end{align*}
where $\theta(p)^\star$ is the true (unadjusted for HIV) logit U5MR in time period $p$ containing the reference time $t$ and $V(t)^{\text{B},\star}$ is the variance.
Note that there are seven estimates, as there are seven five-year age groups of women. We will assume they are independently distributed.

To obtain variance estimates when using the Brass Method we use a technique similar to 
\cite{pedersen:2012} based on the jackknife. We adapt their approach, which is based on deleting clusters, since cluster level information is not available for the census SBH data, and instead define jackknife samples based on women. For the $j$th sample (where the $j$th woman is removed from the dataset), we compute $\hat{\theta}_j(t)^\star= \text{logit}\left( _5\hat{q}_{0,j}(t)^\star\right)$ where $_5\hat{q}_{0,j}(t)^\star$ is computed using the Brass method on this reduced sample. For $n$ total women, we calculate,
\begin{align*}
\bar{\theta}(t)^\star & = \frac{1}{n}\sum_{j=1}^n \hat{\theta}_j(t)^\star, \quad \text{and}\\
\hat{V}(t)^{\text{B},\star} = \widehat{\text{var}}\left(\hat{\theta}(t)^{\text{B},\star}\right) & = \frac{n-1}{n}\sum_{j=1}^n\left\{\hat{\theta}_j(t)^\star - \bar{\theta}(t)^\star\right\}^2,
\end{align*}
where the reference time $t$ is computed using the reference time Brass equation and is held fixed during this procedure. Theoretically, there is uncertainty in the reference time, but as with all previous implementations of Brass we do not consider that here.

\subsection{Combining Brass and Weighted Estimates}

Using FBH data, we can construct weighted (direct) estimates and associated variances for each FBH survey on the 5-year period scale. For survey $surv$ in period $p$, denote the estimate of the logit U5MR as $\hat{\theta}(p;surv)^{\text{W},\star}$ and estimated variance $\hat{V}(p;surv)^{\text{W},\star}$. Again we assume,
\begin{align*}
\hat{\theta}(p;surv)^{\text{W},\star} \sim N\left(\theta(p)^\star, V(p;surv)^{\text{W},\star}\right).
\end{align*}

We follow 
\cite{wakefield:etal:18} and use HIV multiplicative correction factors on the U5MR, obtained using the approach of 
\cite{walker:etal:12} to adjust for HIV bias. This correction is needed to adjust for women who died during the HIV epidemic in Malawi and are therefore not included in the survey. These women would tend to have children with worse outcomes, thus mortality would be biased downwards and the aim of the adjustment terms is to correct for this. They are survey specific and the HIV correction factors are depicted in Figure \ref{Fig:hiv}. For an unadjusted estimate and associated variance, we sample 100,000 realizations and transform them using the HIV correction factors, denoted $k(p;surv)$,
\begin{align*}
\phi^\star(p;surv) & \sim N\left(\hat{\theta}(p;surv)^{\text{W},\star}, \hat{V}(p;surv)^{\text{W},\star}\right),\\
\theta(p;surv) & = \text{expit}\left\{\frac{\phi^\star(p;surv)}{k(p;surv)}\right\}.
\end{align*}
The means and variances of the HIV adjusted  samples are computed and denoted $\hat{\theta}(p;surv)^\text{W},$ and $\hat{V}(p;surv)^\text{W}$, respectively. The same procedure is used for the Brass estimates and variances. 

These corrected estimates and variances are then combined over all surveys and times $t$ that fall in period $p$ via inverse variance weighting. 
Therefore, the overall HIV-corrected estimate and HIV-corrected variance are respectively,
\begin{align*}
\hat{\theta}(p) & = \hat{V}(p)\times \Bigg[\sum_{surv}\left\{\hat{V}(p;surv)^{\text{W}}\right\}^{-1} \hat{\theta}(p;surv)^{\text{W}} + \sum_{\text{t} \in p}\left\{\hat{V}(t)^{\text{B}}\right\}^{-1}\hat{\theta}(t)^{\text{B}}\Bigg],\\
\hat{V}(p) & = \left[ \sum_{surv}\left\{\hat{V}(p;surv)^{\text{W}}\right\}^{-1} + \sum_{\text{t} \in p}\left\{\hat{V}(t)^{\text{B}}\right\}^{-1}\right]^{-1}.
\end{align*}

\section{Simulation Study \label{Sec:simulation}}

\begin{figure}[t]
\centerline{
\includegraphics[width=7.5in]{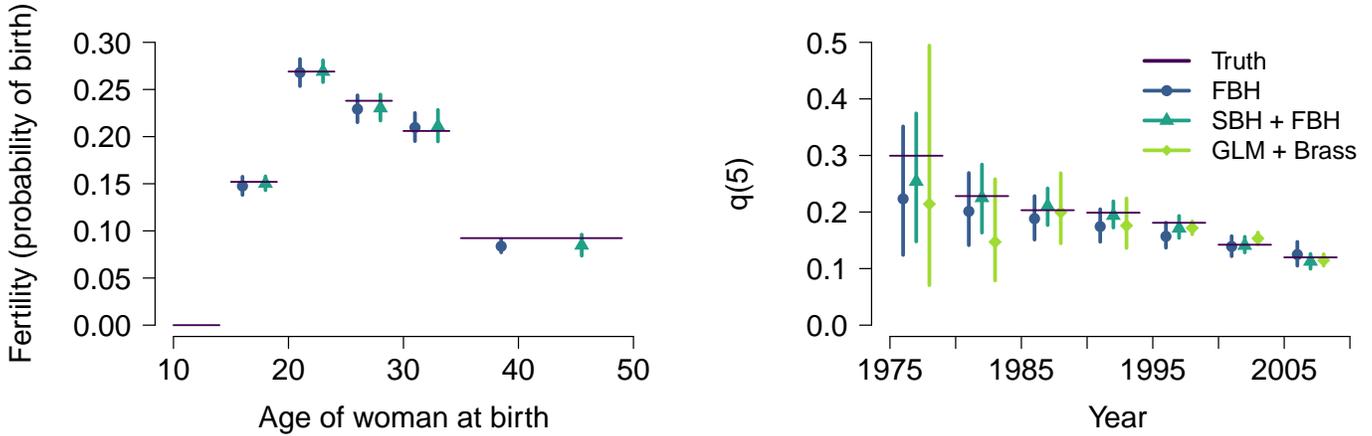}}
\vspace*{-3pt}
\caption{Fertility probabilities (left), 
and U5MR (q(5)) over time (right) used in the simulation. 
Horizontal solid lines indicate the truth. Points indicate posterior medians and vertical lines indicate 95\% credible intervals using only FBH data and both FBH and SBH data using our model (circles and triangles, respectively). Diamonds are estimates of q(5) obtained by combining the Brass method fit on SBH data and a binomial logistic (GLM) fit to FBH data by period.  A color version of this figure can be found in the electronic version of the article.
}
\label{Fig:sim}
\end{figure}

We illustrate the gains that can be made by incorporating SBH data using our approach in a simulation study. We hold fertility fixed over time, but set mortality to vary in 5-year increments. Fertility probabilities and discrete hazards for each period are presented as horizontal lines in Figures \ref{Fig:sim} and \ref{Fig:dh}, respectively. As evident from the figures, fertility was set to be constant over five-year age groups of women and closely resembles fertility patterns observed in the 2010 Malawi DHS 
\citep{MalawiDHS:10}.
Additionally, we use three distinct discrete hazards for each five-year time period with $_1q_0$ and $_1q_a$ for $a=1,\dots,4$ roughly following the North regional model life table.  Birth histories for a total of 5,000 women aged 15 to 49 were simulated. The distribution of ages roughly followed the 2010 Malawi DHS. We will consider two surveys taken in 2010, with one survey containing FBH information for 1,000 women and the other survey containing SBH information for the remaining 4,000 women. More details on simulating the data are provided in Appendix \ref{sec:apb}. 

We take as our model for fertility, $\text{logit}(f(m,\bx(t))) = \beta_{c[m]}$, where the probability of birth depends on the mother's age, via 5 factors,
\begin{align}
c[m] & = \begin{cases}
1 & m=15,\dots,19\\
\vdots & \vdots\\
4 & m=30,\dots,34\\
5 & m=35,\dots,49
\end{cases} \label{eq:c}
\end{align}
and assign independent priors $\beta_{c[m]} \sim \text{N}(0,10^2)$. For mortality, we define 3 age groups
\begin{align}
b[a] = \begin{cases}
0 & a=0\\
1 & a=1,\dots,4\\
2 & a = 5,\dots
\end{cases} \label{eq:b}
\end{align}
and set $\text{logit}(_1q_a(\bx(t))) = \phi_{b[a]}(p) + \beta_{b[a]}$ where $\phi_{b[a]}(p)$ is a temporal random effect that follows a random walk of order 2 (RW2) model that depends on a precision $\kappa$. 
We take the penalized complexity prior for precision as a hyperprior for $\kappa$ 
\citep{simpson:etal:17}.

Several different analyses are considered. First, a complete case analysis is performed (i.e., using the 1,000 women with FBH data). Second, we use our proposed DA approach and include the SBH data on 4,000 women and combine with the FBH data. In both analyses, we use the aforementioned models and use a Hamiltonian Monte Carlo (HMC) algorithm 
\citep{neal:2011} to sample from the posterior during the parameter update step. This approach is explained further for the Malawi example in Appendix \ref{sec:apd}. Finally, we follow Section \ref{Sec:combomethod} and combine the Brass method fit on SBH data with estimates from fitting a logistic regression model to the FBH data; that is,
\begin{align*}
Z_a(t) | _1q_a(t) & \sim \text{Bernoulli}( _1q_a(t)),\\
\text{logit}(_1q_a(t)) & = \beta_{c[a], p},
\end{align*}
where a separate model is fit for each 5-year time period $p$. 

Results are depicted in Figures \ref{Fig:sim} and \ref{Fig:dh}.
Uncertainty from the method that uses Brass is substantially greater in earlier periods. This is because the SBH data only contributes to the 3 most recent time periods. However, this approach tends to produce reasonable estimates in the more recent time periods (diamond points). This is interesting since the Brass method is based on assuming a demographic life model, which may not hold in this simulation. However, since many of the numbers used in the simulation are based on observed fertility and mortality data, this likely explains the good performance of Brass.
We note that, as expected, estimates that incorporate the women with SBH data to the FBH data tend to be better (closer to the truth) and have narrower credible intervals than those obtained from the complete case analysis (see Tables \ref{Tab:fertsim} and \ref{Tab:hazsim}).

\section{Application to Central Malawi}

\label{Sec:results}


\subsection{Model}

Based on exploratory analysis and well known differences in fertility across different covariate groups (National Statistical Office - NSO/Malawi and ICF Macro, 2011), we propose the following model for fertility,
\begin{align}
\text{logit}\left(f(m,\bx(t))\right) & = \phi_{c[m]}(p) + \beta_{m} + \beta_{strata} \label{eq:fert}
\end{align}
where
$\phi_{c[m]}(p)$ is a mother's age group specific RW2 in (roughly) 5-year time periods $p$. We take $c[m]$ to be the same as in the simulation (see Equation (\ref{eq:c})). However, since we observe women in the available FBH data to have given birth between ages $9-48$, we adjust $c[m]$ accordingly: another group for women aged $9-14$ is added and the oldest age group is truncated at age $48$. Since older women are not observed in earlier time periods, women over age 35 are grouped together for the RW2 model. There are 11 time periods $p$: $1964-1969, 1970-1974, 1975-1979, \dots, 2010-2014, 2015-2019$. We include a fixed effect for each age $m$ (ages $9-11$ were grouped together because of many 0 counts). Thus, women in age groups defined by $c[m]$ have the same trend in fertilities, but different overall levels of fertility by age. We also include a fixed effect for strata (urban and rural).


For the mortality model, we will assume for simplicity that the probability of death within one year is the same for children ages 5 and older. We propose the following model for mortality,
\begin{align}
Z_a(t)\ |\ _1q_a^\star(\bx(t)) & \sim \text{Bernoulli}(_1q_a^\star(\bx(t)))\nonumber \\
_1q_a^\star(\bx(t)) &=  k(t;surv) \times {_1}q_a(\bx(t)) \nonumber \\
\text{logit}\left(_1q_{a}(\bx(t))\right) & = \beta_{SBH,strata} + \beta(c[a],\bx(t))\nonumber \\
\beta(c[a],\bx(t)) & = \phi_{b[a]}(t) + \beta_{c[a]} + \beta_{district} + \beta_{strata} \label{eq:haz}
\end{align}
where $k(t;surv)$ is a term for HIV bias that varies by year $t$ and depends on the year the survey was conducted. The term $\beta_{SBH,strata}$ allows for bias in SBH (census data) by urban/rural. In an exploratory analysis, we found that women living in rural areas in the census reported a higher proportion of their children dead than similarly aged women in the surveys taken either side of that time period (see Figure \ref{Fig:eda1}). When making predictions, we do not include the HIV or SBH bias terms. We include a RW2 in years, $\phi_{b[a]}(t)$, by age group $b[a]$ (see Equation (\ref{eq:b})).
We include fixed effects for strata and district and also for age group with
\begin{align*}
c[a] & = \begin{cases}
0 & a=0\\
1 & a=1\\
\vdots & \vdots \\
4 & a=4\\
5 & a=5,\dots
\end{cases}.
\end{align*} 
Therefore, the trend in logit hazards is the same for ages 1--4, but each age has its own level of mortality.

\subsection{Computation}
In the simulation, we had supposed that women could not give birth in the year of the survey, thus a woman's fertility history and child's life trajectory are fully observed up to the time of the survey (see Appendix \ref{sec:apb}). Clearly, this is not a reasonable simplification for applying our method to the Malawi data. We describe the special considerations regarding the year of the survey in Appendix \ref{sec:apc}. Briefly, because we are working on a discrete time scale, it is possible for women to give birth during the year of the survey. We accommodate this and acknowledge the shortened observation time by creating adjustment terms for the fertilities and hazards.

We again implement an HMC algorithm for the parameter update step. Forms for the negative log posteriors and gradients along with hyperpriors are given in Appendix \ref{sec:apd}.

\subsection{Results}
We fit four different models to the data. First, we computed weighted estimates based only on FBH data. Second, we used the approach described in Section \ref{Sec:combomethod} to combine the direct estimates with indirect estimates obtained using the Brass method at the district-level on SBH data. Both of these were done on the period (5-year time scale) since there were insufficient data to support separate hazards on the yearly time scale. Third, we fitted (\ref{eq:fert}) and (\ref{eq:haz}) to all available FBH data (DHS and MICS). Finally, we used our DA approach to incorporate SBH data and fitted models (\ref{eq:fert}) and (\ref{eq:haz}) to all data. Trace plots for the mortality model are in Figure \ref{Fig:tracehaz}.

\begin{figure}[!t]
\centerline{\includegraphics[width=6.5in]{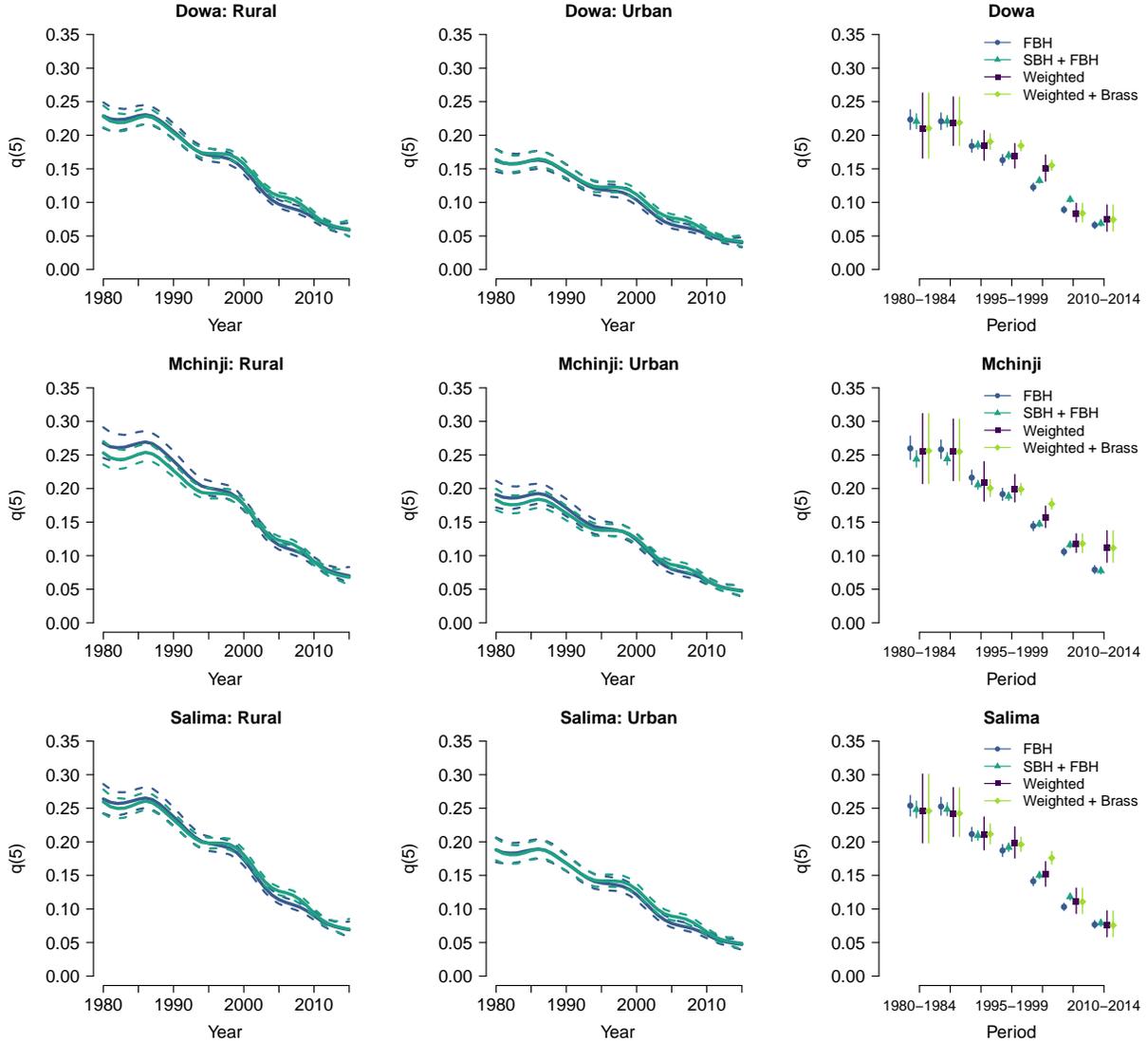}}
\caption{Left and middle panel: posterior medians (points and solid lines) and 95\% credible intervals (dashed lines) for U5MR ($q(5)$) in 3 districts. Right panel: estimates of U5MR (points) and 95\% uncertainty interval (lines).  A color version of this figure can be found in the electronic version of the article.}
\vspace*{-3pt}
\label{Fig:malawi}
\end{figure}

Results for fertility are found in Figures \ref{Fig:fertr}--\ref{Fig:fert} and show similar results between FBH only and FBH and SBH analyses. 
Figure \ref{Fig:malawi} (left and middle panels) compares posterior medians and 95\% credible intervals for the U5MR in three districts for illustration (results for the other districts are in Figures \ref{Fig:q5a}--\ref{Fig:q5b}). The posteriors tend to be similar when we add in the census data and use our proposed model. We aggregate to five year time periods in order to compare with the other methods. To do this, we sample births using the fertility samples for our FBH only and FBH and SBH models. We then use them to average the samples of $q(5)$ over strata and years within 5-year time periods. More details are in Appendix \ref{sec:ape}.

In the right panel of Figure \ref{Fig:malawi} and Figures \ref{Fig:q5a}--\ref{Fig:q5b}, we see that the indirect estimates from Brass contribute to 3 time periods: 1990--1994, 1995--1999, and 2000--2004 and often result in higher $q(5)$. In general, the estimates are similar across the other three methods. Uncertainty is much higher when using the weighted estimates on FBH data as compared to using our proposed model.  

To assess the accuracy of the four models, we divide the FBH data into training and holdout data. We split the 2006 MICS and 2010 DHS into two roughly equal-sized groups: a training and validation set. The validation set consisted of 331 clusters and the remaining 314 were used in the training set  (see Table \ref{Tab:data_descrip}). We refit the models to the training data (2004 DHS, 2008 Census, 2013 MICS, 2015 DHS, and training sets from the 2006 MICS and 2010 DHS) and assess the accuracy of estimation using a weighted mean squared error (MSE). Denote the estimate of logit of U5MR in district $d$ time period $p$ for model $M$, $Y_{dp}^{(M)}$. For the weighted and weighted and Brass estimates, we use the asymptotic variance of the estimate and for the Bayesian analyses, the posterior variances.  These four estimates and variances are compared with the weighted estimates of the logit of U5MR from the holdout clusters, $y_{dp}$ taken to be the truth. We assess the accuracy by time period over all $9$ districts by using a weighted MSE:
\begin{align*}
\text{MSE}(p)^{(M)} 
=  \sum_{d=1}^9 w_{dp} \left\{E\left(Y_{dp}^{(M)} - y_{dp}\right)\right\}^2  + \sum_{d=1}^9 w_{dp} \text{Var} \left(Y_{dp}^{(M)}\right)
\end{align*}
where $p = \{1985-1989, 1990-1994, 2000-2004, 2005-2009\}$ and $w_{dp} = \hat{V}_d(p)^{-1}/\sum_{d=1}^9 \hat{V}_d(p)^{-1}$, with $\hat{V}_d(p)^{-1}$ denoting the variance of the weighted estimates in district $d$ and time period $p$. This allows us to upweight the MSE in districts and periods where the ``truth'' is more certain.

Models involving direct estimates tend to perform worse (Table \ref{Tab:mse} and Figure \ref{Fig:mse}). Adding SBH via the Brass method or the DA approach in general improves the predictions. In our smoothed model, the variance component of the MSE is $55\%$ lower when SBH data are included; however, the bias tends to be higher. In Appendix \ref{sec:apf} we carry out another comparison based on the percentage absolute relative error (PARE) of the probabilities $q(5)$. The substantive results are changed slightly with the DA procedure producing a PARE of 8.8\%, compared to 9.0\% for the FBH analysis only, the direct estimate PARE was 10.5\% and the direct and indirect combined was 11.1\%. Figure \ref{Fig:pare} and Table \ref{Tab:pare} contain the PARE values by period.

\begin{table}[h]
\caption{Mean Squared Error $\times 100$ by Model.}
\label{Tab:mse}
\begin{center}
\begin{tabular}{lrrrr}
\hline
Period & Weighted Estimates & Weighted Estimates  & Smoothed Model: & Smoothed Model: \\
& & + Brass & FBH & FBH + SBH \\
\hline
1985--1989 & 3.84 & 3.83 & 2.69 & 4.13 \\  
1990--1994 & 2.65 & 2.08 & 1.86 & 2.04 \\
1995--1999 & 2.92 & 1.64 & 1.56 & 1.78 \\  
2000--2004 & 2.13 & 2.62 & 1.21 & 0.33 \\  
2005--2009 & 5.83 & 5.82 & 5.10 & 4.19 \\
Average & 3.07 & 2.79 & 2.06 & 1.99 \\
\hline
\end{tabular}
\end{center}
\end{table}

\section{Discussion}
\label{Sec:conclusion}

We have presented a novel framework for analyzing SBH data in the context of U5MR estimation. Using our approach, data from a variety of sources that contain birth history information at varying degrees of detail can be analyzed together to produce an estimate of U5MR over time with a measure of  uncertainty. Our method falls under the umbrella of data augmentation, where we impute FBH information for the SBH data.

We make some simplifying assumptions since we opt to work on a discrete time scale; however, modifications to our approach can be made. For example, multiple births and having a lag time between births could be accommodated by adjusting equation (\ref{eq:substep1}) and the model for fertility. Further, the approach could be extended to a monthly time scale or continuous time. We also define the fertility patterns by the age of the women rather than time since fist birth or marriage; however, both of these variations are possible with our framework.

In our application, we included fixed effects for strata and districts and did not have these vary in time. One future consideration is using information on migration and allowing the district of the children and women to change over time. 
Other terms, such as a spatial random effect could be included.
Additionally, terms for mother's age, number of births, and if available, sex of the child, which may be informative of U5MR, could also be included. However, producing an overall estimate of U5MR at an administrative level would be more complicated since the estimate would need to be marginalized over the age, birth number, and sex population distributions. 
For understanding individual-level associations, including these terms into the hazard model would be useful and straightforward to implement in our approach.

In our simulation, we found that by incorporating SBH data uncertainty is reduced. When we applied our method to actual data, adjustments to our model were made to accommodate differences between the SBH census data and FBH survey data. In general, we assumed that FBH data were the ``gold standard'' and adapted our model to reflect this. A fixed effect term for SBH data by strata was added into the mortality model to adjust for the proportion of deaths as reported by SBH women being higher in rural areas than expected (compared to FBH women). We also used a portion of the FBH data as holdout data to validate the models. It is not immediately clear what the truth is in this scenario, however. There are well known biases in both types of data, making model assessment difficult 
\citep{silva:2012}. For SBHs, women may omit live births since they are not asked to systematically recall every birth. In the DHS and MICS, women are asked detailed questions about all of their children (including those that no longer live with them) to limit this type of bias. In FBHs, women may misreport when births and deaths occurred. Furthermore, additional questions on pregnancy and postnatal care are asked if a woman has had recent births, thus displacement of births or omission of recent births may occur.
Special consideration for the data sources and outcome of interest is needed when developing the fertility and hazard models, and this is highly context specific.

\newpage

 \bibliographystyle{apalike} 
 \bibliography{arxiv.bib}

\begin{thebibliography}{}

\bibitem[Alkema and New, 2012]{alkema2012progress}
Alkema, L. and New, J.~R. (2012).
\newblock Progress toward global reduction in under-five mortality: a bootstrap
  analysis of uncertainty in millennium development goal 4 estimates.
\newblock {\em PLoS Medicine}, 9(12):e1001355.

\bibitem[Alkema and New, 2014]{alkema:2014b}
Alkema, L. and New, J.~R. (2014).
\newblock Global estimation of child mortality using a bayesian b-spline
  bias-reduction model.
\newblock {\em The Annals of Applied Statistics}, pages 2122--2149.

\bibitem[Alkema et~al., 2014]{alkema:2014a}
Alkema, L., New, J.~R., Pedersen, J., You, D., et~al. (2014).
\newblock Child mortality estimation 2013: an overview of updates in estimation
  methods by the {U}nited {N}ations {I}nter-agency {G}roup for {C}hild
  {M}ortality {E}stimation.
\newblock {\em PLoS ONE}, 9(7):e101112.

\bibitem[Brady and Hill, 2017]{brady:2017}
Brady, E. and Hill, K. (2017).
\newblock Testing survey-based methods for rapid monitoring of child mortality,
  with implications for summary birth history data.
\newblock {\em PLoS ONE}, 12(4):e0176366.

\bibitem[Brass, 1964]{brass:64}
Brass, W. (1964).
\newblock {\em Uses of census or survey data for the estimation of vital
  rates}.
\newblock United Nations.
\newblock Paper prepared for the African Seminar on Vital Statistics, Addis
  Ababa, 14--19 December, 1964.

\bibitem[Coale and Demeny, 1983]{coale:demeny:83}
Coale, A.~J. and Demeny, P. with~Vaughan, B. (1983).
\newblock {\em Regional Model Life Tables and Stable Populations}.
\newblock New York: Academic Press, 2nd edition.

\bibitem[Coale and Trussell, 1977]{coale:trussell:77}
Coale, A.~J. and Trussell, J. (1977).
\newblock Annex {I}: estimating the time to which {B}rass estimates apply.
\newblock {\em Population Bulletin of the United Nations}, 10:87--89.

\bibitem[Feeney, 1980]{feeney:80}
Feeney, G. (1980).
\newblock Estimating infant mortality trends from child survivorship data.
\newblock {\em Population Studies}, 34(1):109--128.

\bibitem[Golding et~al., 2017]{golding:etal:17}
Golding, N., Burstein, R., Longbottom, J., Browne, A.~J., Fullman, N.,
  Osgood-Zimmerman, A., et~al. (2017).
\newblock Mapping under-5 and neonatal mortality in {A}frica, 2000--15: a
  baseline analysis for the {S}ustainable {D}evelopment {G}oals.
\newblock {\em The Lancet}, 390(10108):2171--2182.

\bibitem[Hill, 2013]{hill:2013}
Hill, K. (2013).
\newblock Indirect estimation of child mortality.
\newblock In Moultrie, T.~A., Dorrington, R., Hill, A., Hill, K., Tim{\ae}us,
  I., and Zaba, B., editors, {\em Tools for Demographic Estimation}, pages
  148--164. International Union for the Scientific Study of Population Paris.

\bibitem[Hill et~al., 2015]{hill:15}
Hill, K., Brady, E., Zimmerman, L., Montana, L., Silva, R., and Amouzou, A.
  (2015).
\newblock Monitoring change in child mortality through household surveys.
\newblock {\em PLoS ONE}, 10(11):e0137713.

\bibitem[Hill and Figueroa, 1999]{hill:99}
Hill, K. and Figueroa, M.-E. (1999).
\newblock Child mortality estimation by time since first birth.
\newblock In Zaba, B. and Blacker, J., editors, {\em Brass Tacks: Essays in
  Medical Demography}, pages 9--19. London: Athlone.

\bibitem[Hill et~al., 2012]{hill:12}
Hill, K., You, D., Inoue, M., and Oestergaard, M.~Z. (2012).
\newblock Child mortality estimation: accelerated progress in reducing global
  child mortality, 1990--2010.
\newblock {\em PLoS Medicine}, 9(8):e1001303.

\bibitem[Horvitz and Thompson, 1952]{horvitz:thompson:52}
Horvitz, D. and Thompson, D. (1952).
\newblock A generalization of sampling without replacement from a finite
  universe.
\newblock {\em Journal of the American Statistical Association}, 47:663--685.

\bibitem[Mercer et~al., 2015]{mercer2015space}
Mercer, L.~D., Wakefield, J., Pantazis, A., Lutambi, A.~M., Masanja, H., and
  Clark, S. (2015).
\newblock Space--time smoothing of complex survey data: Small area estimation
  for child mortality.
\newblock {\em The Annals of Applied Statistics}, 9(4):1889--1905.

\bibitem[{National Statistical Office - NSO/Malawi and ICF Macro},
  2011]{MalawiDHS:10}
{National Statistical Office - NSO/Malawi and ICF Macro} (2011).
\newblock Malawi {D}emographic and {H}ealth {S}urvey 2010.
\newblock Final report, NSO/Malawi and ICF Macro, Zomba, Malawi.
\newblock Available at http://dhsprogram.com/pubs/pdf/FR247/FR247.pdf.

\bibitem[Neal, 2011]{neal:2011}
Neal, R. (2011).
\newblock {MCMC} using {H}amiltonian dynamics.
\newblock In Brooks, S., Gelman, A., Jones, G., and Meng, X., editors, {\em
  Handbook of Markov Chain Monte Carlo}, volume~2, pages 113--162. Chapman and
  Hall/CRC Press.

\bibitem[Pedersen and Liu, 2012]{pedersen:2012}
Pedersen, J. and Liu, J. (2012).
\newblock Child mortality estimation: appropriate time periods for child
  mortality estimates from full birth histories.
\newblock {\em PLoS Medicine}, 9(8):e1001289.

\bibitem[Pezzulo et~al., 2016]{pezzulo:etal:16}
Pezzulo, C., T.Bird, Edson, C., Utazi, C., Sorichetta, A., Tatem, A.,
  Yourkavitch, J., and Burgert-Brucker, C. (2016).
\newblock Geospatial modeling of child mortality across 27 countries in
  sub-{S}aharan {A}frica.
\newblock Technical report, ICF International.
\newblock DHS Spatial Analysis Reports No. 13.

\bibitem[Preston et~al., 2000]{preston2000demography}
Preston, S.~H., Heuveline, P., and Guillot, M. (2000).
\newblock {\em Demography: Measuring and Modeling Population Processes}.
\newblock Blackwell Malden, MA.

\bibitem[Rajaratnam et~al., 2010]{rajaratnam:2010}
Rajaratnam, J.~K., Tran, L.~N., Lopez, A.~D., and Murray, C.~J. (2010).
\newblock Measuring under-five mortality: validation of new low-cost methods.
\newblock {\em PLoS Medicine}, 7(4):e1000253.

\bibitem[Rutstein and Rojas, 2006]{rutstein:06}
Rutstein, S.~O. and Rojas, G. (2006).
\newblock {\em Guide to {DHS} statistics}.
\newblock Calverton, MD: ORC Macro.

\bibitem[Silva, 2012]{silva:2012}
Silva, R. (2012).
\newblock Child mortality estimation: consistency of under-five mortality rate
  estimates using full birth histories and summary birth histories.
\newblock {\em PLoS Medicine}, 9(8):e1001296.

\bibitem[Simpson et~al., 2017]{simpson:etal:17}
Simpson, D., Rue, H., Riebler, A., Martins, T., and S{\o}rbye, S. (2017).
\newblock Penalising model component complexity: a principled, practical
  approach to constructing priors (with discussion).
\newblock {\em Statistical Science}, 32:1--28.

\bibitem[S{\o}rbye and Rue, 2014]{sorbye:rue:14}
S{\o}rbye, S. and Rue, H. (2014).
\newblock Scaling intrinsic {G}aussian {M}arkov random field priors in spatial
  modelling.
\newblock {\em Spatial Statistics}, 8:39--51.

\bibitem[Trussell, 1975]{trussell:75}
Trussell, J. (1975).
\newblock A re-estimation of the multiplying factors for the {B}rass technique
  for determining childhood survivorship rates.
\newblock {\em Population Studies}, 29:97--107.

\bibitem[{United Nations}, 1983]{manualX:83}
{United Nations} (1983).
\newblock {\em {M}anual {X}: Indirect Techniques for Demographic Estimation}.
\newblock United Nations, New York.

\bibitem[Wakefield et~al., 2018]{wakefield:etal:18}
Wakefield, J., Fuglstad, G.-A., Riebler, A., Godwin, J., Wilson, K., and Clark,
  S.~J. (2018).
\newblock Estimating under five mortality in space and time in a developing
  world context.
\newblock {\em Statistical Methods in Medical Research}.
\newblock To Appear.

\bibitem[Walker et~al., 2012]{walker:etal:12}
Walker, N., Hill, K., and Zhao, F. (2012).
\newblock Child mortality estimation: methods used to adjust for bias due to
  {AIDS} in estimating trends in under-five mortality.
\newblock {\em PLoS Medicine}, 9:e1001298.

\end{thebibliography}
 
\newpage

\appendix

\counterwithin{figure}{section}
\counterwithin{table}{section}

\section{Descriptives for Malawi Application}
\label{sec:apa}
Table \ref{Tab:data_descrip} contains the number of clusters, women, births, deaths by survey type. 

\begin{table}[b]
\caption{Data summaries, by survey.}
\label{Tab:data_descrip}
\begin{center}
\begin{tabular}{llrrrr}
\hline
\multicolumn{2}{l}{Survey} & No. Clusters & No. Women & No. Births & No. Deaths \\
\hline
Census 2008 & & & 71,618 & 332,055 & 62,884 \\
DHS 2004 & & 186 & 4,199 & 13,394 & 2,532\\
\multirow{2}{*}{DHS 2010} & Training  & 140 & 3,879 & 12,626 & 2,125 \\
& Holdout & 145 & 3,983 & 12,873 & 2,066 \\
DHS 2015 & & 284 & 8,417 & 23,240 & 2,621\\
\multirow{2}{*}{MICS 2006} & Training & 174 & 4,518 & 14,505 & 2,448 \\
& Holdout & 186 & 4,832 & 15,069 & 2,397 \\
MICS 2013 & & 381 & 8,267 & 24,798  & 3,028\\
\hline
\end{tabular}
\end{center}
\end{table}

\begin{figure}[b]
\includegraphics[width=\linewidth]{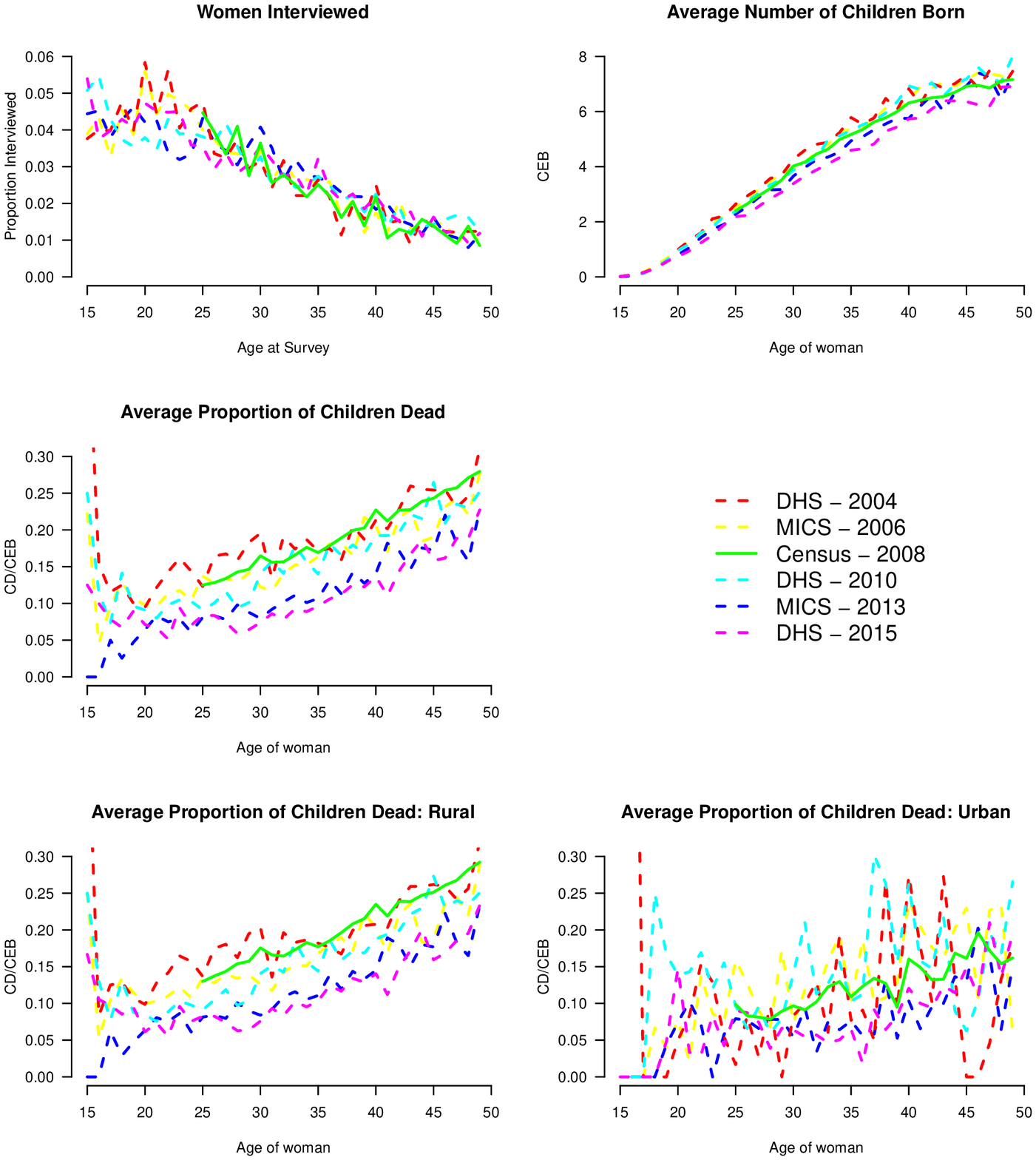}
  \caption{Top Left: Proportion of women between ages 15--49 interviewed. Top Right: Average number of reported children ever born (CEB). Middle Left: Average number of children dead (CD) to CEB. Bottom Row: Average number of CD to CEB by rural and urban strata.}
\vspace*{-3pt}
\label{Fig:eda1}
\end{figure}

Results from an exploratory analysis, similar to the diagnostic measures reported by \cite{hill:15}, where FBH information was transformed to SBH information are depicted in Figure \ref{Fig:eda1}. The proportion of women interviewed tends to be similar across surveys. There appears to be some age heaping present, where women report their age as ending in ``5'' or ``0.'' There also do not appear to be systematic differences between the number of total births women report having had across surveys. However, there appear to be differences in the ratio of children dead (CD) to children ever born (CEB). Importantly, the differences tend to vary in rural and urban areas. We would expect the proportion CD among women in the census to be similar to the proportion CD among women in the 2006 MICS and 2010 DHS since these surveys were taken closest in time. These surveys are much noisier, but we can see that in rural areas the proportion CD is higher than both the 2006 MICS and 2010 DHS across almost all women. This is not the case in urban areas. This trend motivates including a term in our model that can capture this pattern.

\begin{figure}[b]
\includegraphics[width=\linewidth]{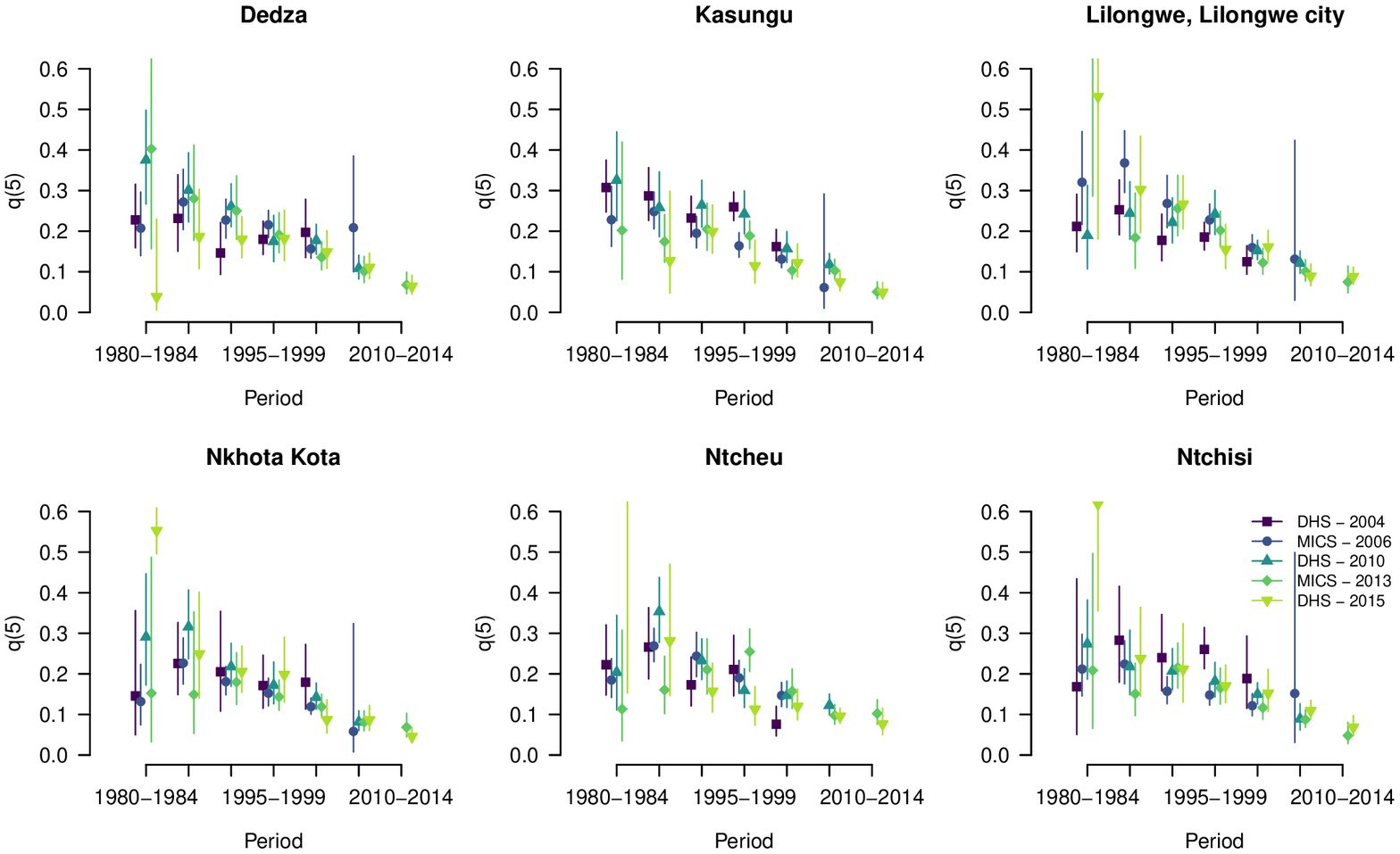}
  \caption{Direct estimates by survey for 6 districts. Points are the estimates and lines are the 95\% Wald-based confidence intervals computed on the logit scale.}
\vspace*{-3pt}
\label{Fig:supp-de}
\end{figure}

\begin{figure}[b]
  \centerline{\includegraphics[width=6in]{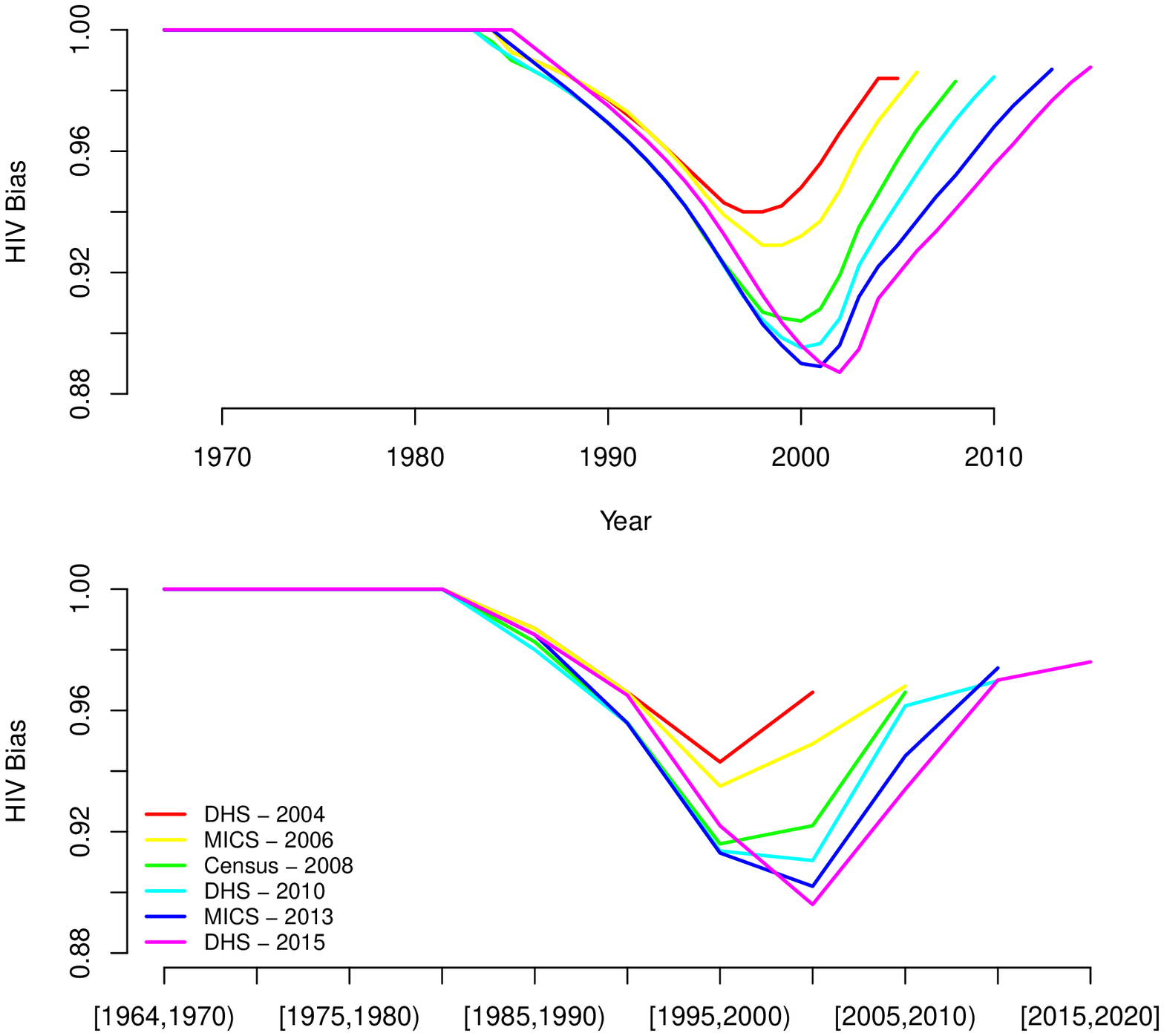}}
  \caption{Top: yearly HIV bias color-coded by survey in Malawi. Bottom: 5-year HIV bias color-coded by survey in Malawi.}
\vspace*{-3pt}
\label{Fig:hiv}
\end{figure}

Figure \ref{Fig:supp-de} shows the HIV-adjusted direct estimates by district. Figure \ref{Fig:hiv} shows the HIV bias as a function of survey. HIV bias factors were computed following \cite{walker:etal:12}.

\begin{figure}[b]
\includegraphics[width=5.5in]{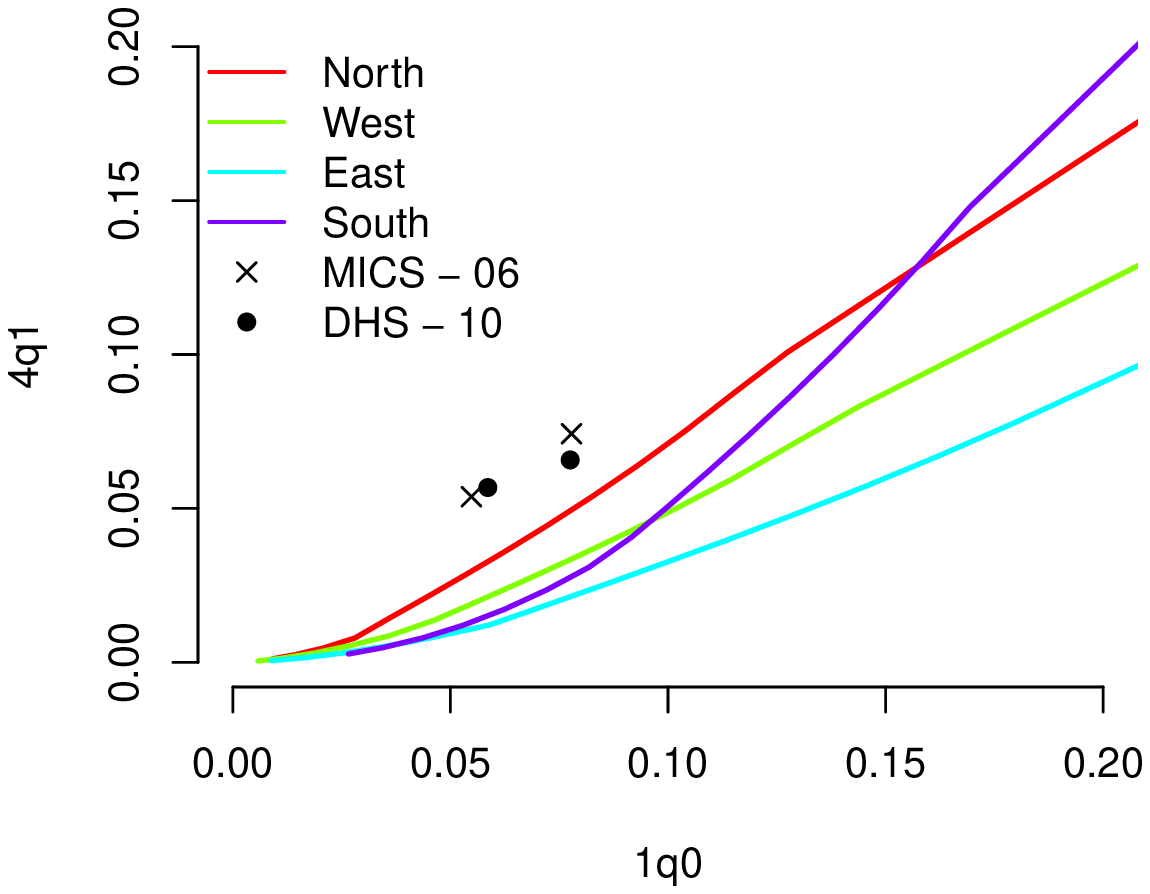}
  \caption{Direct estimates of $_1q_0$ and $_4q_1$ obtained from 2006 MICS and 2010 DHS for the time periods 2000--2004 and 2005--2009 compared to the 4 Coale-Demeny regional model life tables.}
\vspace*{-3pt}
\label{Fig:brass-models}
\end{figure}

To determine which model life table to use in analyses involving the Brass method, we computed the direct estimates of $_1q_0$ and $_4q_1$ for the 2006 MICS and 2010 DHS and compared them to the 4 Coale-Demeny regional models (Figure \ref{Fig:brass-models}) as suggested by \cite{hill:2013} . Although none of the models fit the observed values exactly, the North model appeared to fit best.

\section{Simulation Details}
\label{sec:apb}

Birth histories for 5,000 women were simulated on a discrete, yearly time scale. For simplicity, we allowed the year prior to the survey to be completely observed and did not allow for births during the survey year, which follows the simple example provided in Section 2.1. Thus, children could be born at any point prior to and including $t_{surv} -1$ (when the woman was aged $m_{surv} -1$), and could die in $t_{surv}$ where $t_{surv}$ is the year of the survey and $m_{surv}$ is the age of the woman at time of survey.

\begin{figure}[b]
  \centerline{\includegraphics[width=5.75in]{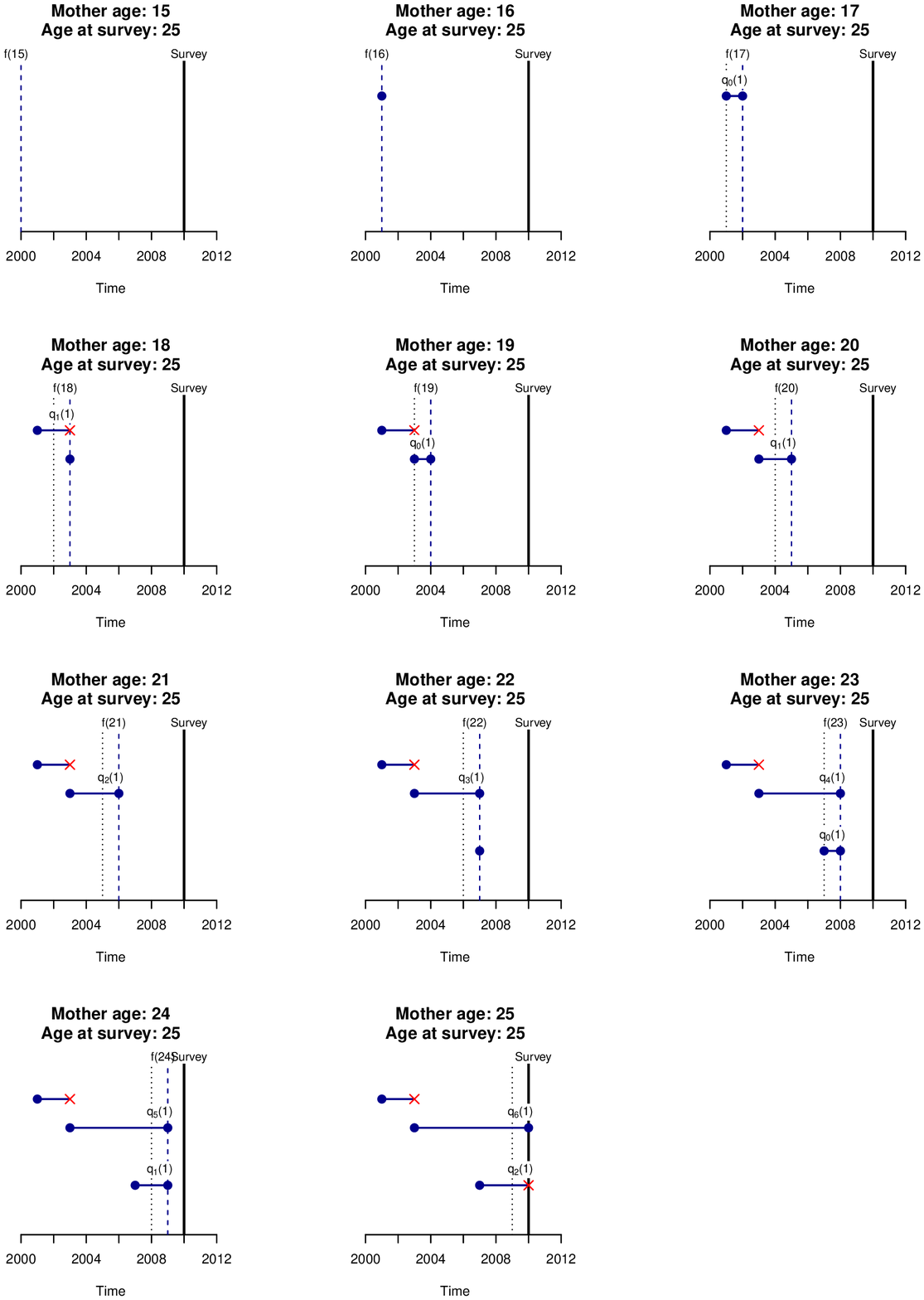}}
  \caption{Illustration of the data generating mechanism along with relevant probabilities. Suppose a woman is 25 at the time of the survey in 2010 and suppose $f(m)>0$ for $m\geq15$. Starting at the top right and proceeding left and down are panels following her and any children she has forward through time starting at age 15. The blue dashed line represents the current year and black dotted line represents the prior year. Blue circles represent births and survival, red ``x''s represent deaths.}
\vspace*{-3pt}
\label{Fig:datagen}
\end{figure}

Figure \ref{Fig:datagen} 
illustrates how FBH data were simulated for a woman who was $25$ at the time of the survey. In the top left panel, when the woman is $15$, the probability she gives birth is $f(15)$ (fertility does not change over time). In this example, she does not give birth. In the middle top panel, when the woman is $16$, the probability she gives birth is $f(16)$. Here, she does give birth. In the following year (top right panel), the probability the woman gives birth is $f(17)$, and the probability the child dies is $_1q_0 = q_0(1)$. As the woman and her children age, we observe her to have 3 children at ages 16, 18, and 23. One child dies between age 1 and 2 and another dies between age 2 and 3. Her other child survives through the time of the survey. For women with FBH data, this information is completely observed. For women with SBH data, we only observe the total number of children the woman had and the number of those children that died (in this example, 3 births and 2 deaths).

\begin{figure}[b]
  \centerline{\includegraphics[width=6in]{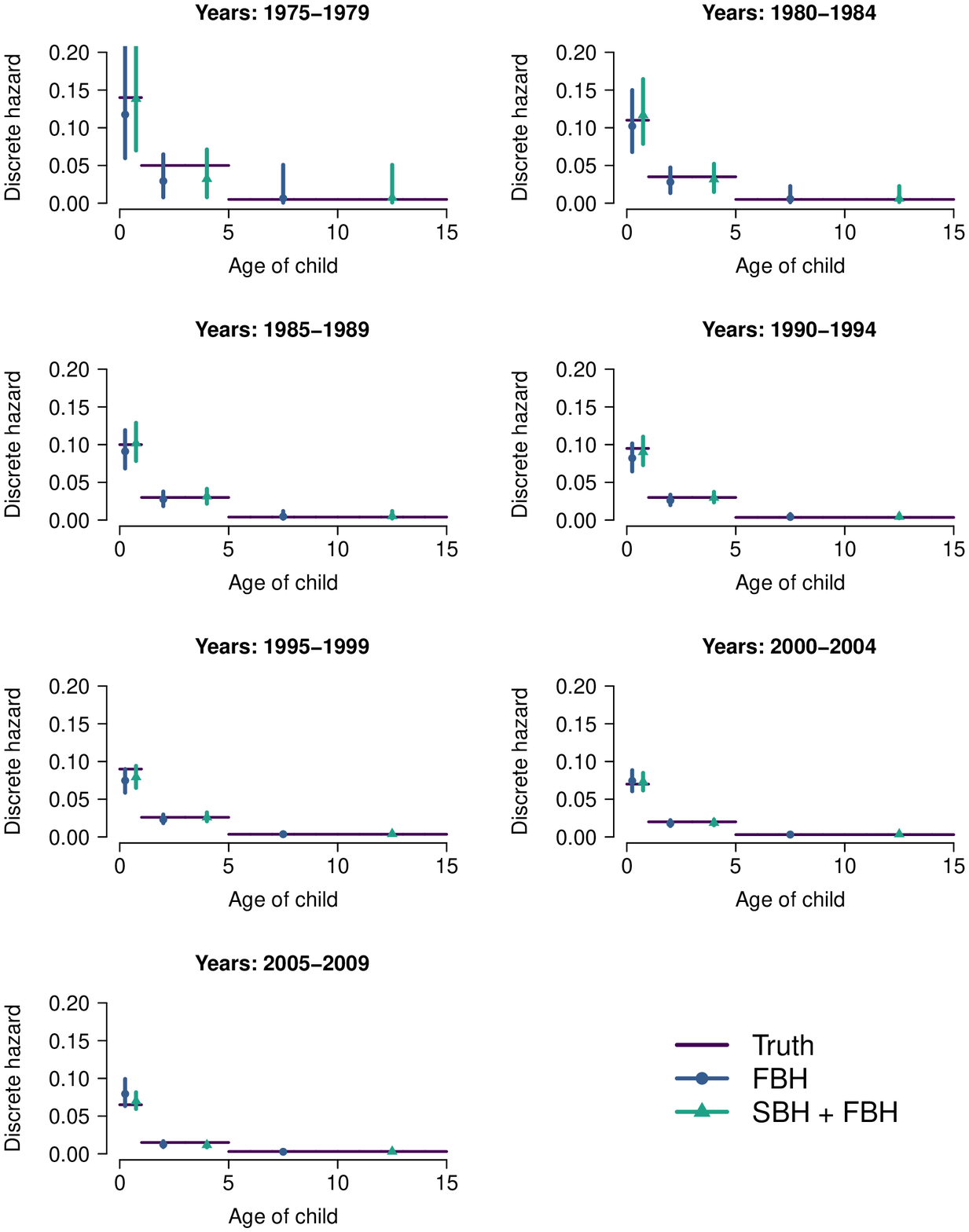}}
\caption{Discrete hazards by time period. 
Note: the probability of death within one year after age 5 was considered to be constant (only up to age 15 is plotted). 
Horizontal solid lines indicate the truth. Points indicate posterior medians and vertical lines indicate 95\% CIs using only FBH data and both FBH and SBH data.
}
\label{Fig:dh}
\end{figure}

Discrete hazards used in the simulation are in Figure \ref{Fig:dh} (solid lines). Results from fitting the model to FBH data only and all data are also shown. The posterior medians are similar and close to the truth and uncertainty is reduced when we incorporate SBH data, as expected.

To assess accuracy we use $\text{Absolute Bias} = |\hat{\psi}^{(M)} - \psi|$ where $\psi$ is the truth for a generic parameter (i.e., $f(m)$ or $ _5q_0(t)$) and $\hat{\psi}^{(M)}$ is an estimate of $\psi$ (e.g. posterior median) obtained from approach/model $M$. To compare measures of uncertainty we take the ratio of the 95\% credible/confidence intervals (CIs). Results are in Tables \ref{Tab:fertsim} and \ref{Tab:hazsim}. In general, absolute bias and the width of the uncertainty intervals is lower when we include SBH data.

\begin{table}[b]
\caption{Summary measures for fertilities, in the simulation. Ratio of CIs are relative to the width of the FBH intervals.}
\label{Tab:fertsim}
\begin{center}
\begin{tabular}{lrrrrr}
\hline
 & \multicolumn{2}{c}{Absolute Bias $\times 100$} & \multicolumn{2}{c}{Width of CIs $\times 100$} & \multicolumn{1}{c}{Ratio of CIs}  \\
 & FBH & SBH + FBH & FBH & SBH + FBH & \\
\hline
15--19 & 0.44 & 0.16 & 1.9 &  1.5 & 0.77 \\
20--24 & 0.10 & 0.015 & 2.9 & 2.3 & 0.81 \\
25--29 & 0.87 & 0.76 & 2.9 & 2.8 & 0.97 \\
30--34 & 0.38 & 0.50 & 3.0 & 3.4 & 1.12 \\
35--49 & 0.85 & 0.76 & 1.4 & 2.3 & 1.59 \\
\hline
\end{tabular}
\end{center}
\end{table}

\begin{table}[b]
\caption{Summary measures for hazards, in the simulation. Ratio of CIs are relative to the width of the FBH intervals.}
\label{Tab:hazsim}
\begin{center}
\begin{tabular}{lrrrrrr}
\hline
 & \multicolumn{3}{c}{Absolute Bias $\times 100$} & \multicolumn{3}{c}{Width of CIs $\times 100$} (Ratio of CIs)  \\
 & FBH & SBH + FBH & GLM + Brass & 
   FBH & SBH + FBH & GLM + Brass \\
\hline
1975--79 & 7.3 & 4.3 & 8.5 & 22 & 23 (1.0) & 42 (1.9)\\
1980--84 & 2.6 & 0.29 & 8.1 & 13 & 12 (0.94) & 18 (1.4)\\
1985--89 & 1.5 & 0.72 & 0.37 & 7.7 & 6.5 (0.84) & 12 (1.6)\\
1990--94 & 2.4 &  0.45 & 2.3 & 5.8 & 4.6 (0.81) & 8.7 (1.5)\\
1995--99 & 2.3 & 0.93 & 0.92 & 4.4 & 3.9 (0.88) & 2.3 (0.52)\\
2000--04 & 0.29 & 0.088 & 1.1 & 3.6 & 2.8 (0.78) & 2.1 (0.58) \\
2005--09 & 0.54 & 0.70 & 0.50 & 4.2 & 2.6 (0.62) & 2.1 (0.49)\\
\hline
\end{tabular}
\end{center}
\end{table}

\section{Date Setup for Malawi Application}
\label{sec:apc}

Calendar time is first defined on yearly scales, based on the month and year of the survey. In order to more accurately align with the woman's age and to reduce the number of corrections (as described later), we divide time into the most recent 6 months prior to the survey and 12 month intervals preceding that. For example, if a survey was taken in February 2009, the time intervals would look like the following:
\begin{itemize}
\item September - December 2008, January and February 2009
\item September - December 2007, January - August 2008
\item ...
\end{itemize}
Based on the month and year reported for the birth of any children, we determine in what time interval the child was born.

The next step is to align the age of the woman during each interval. This is based on assuming that, on average, the woman will be the age in years reported plus six months. For example, if a woman reports that she is 35 at the time of survey, on average she will be 35.5. Thus, the most recent interval will correspond to when she is 35, second most recent when she was 34, etc.

Finally, we discretize the intervals by assigning them a year according to the year of survey. The most recent time period would correspond to the year of the survey and each subsequent interval would be one year prior. Since we smooth over time, the misalignment between the time intervals and assigned year is unlikely to be a major issue. We note that if the survey was taken in June there would be no misalignment.

We will need to make corrections to the model based on the yearly discretization of time. Suppose we are in the most recent time interval (which corresponds to the year of the survey). Since this interval only corresponds to 6 months, the fertility, or probability of birth should be, approximately half of what it normally is. This is assuming constant fertility throughout the year. Mathematically, if $m = m_{surv}$ and $t=t_{surv}$ then $f(m_{surv}, t_{surv}) = 0.5 f(m, t)$. All other years will not require this correction.

Finally, we need to consider a hazard correction. Suppose a child is born in the most recent time period. At most, they have 6 months of exposure time. On average, we might expect them to be born at the midpoint of that time interval, meaning we are interested in the probability they die within 3 months. Clearly, this is less than the probability of dying within 12 months, but the factor is likely to be $>1/4$ since children are at highest risk in the first month of life. We base the correction on what is observed in the FBH data and use $0.65$.

Now, consider a child born in the second most recent interval. On average, the child will be born halfway through the interval. Since the last interval is only 6 months, their exposure time is, on average, one year. We will not use any adjustments for the earlier periods. However, we note that for a small portion of women, they have children that, in terms of the newly defined discretized time, die after the survey. For example, consider a child born 1.5 years before the survey, which using our discretized intervals is 1 year before the survey. In our setup, they are only at risk for one year, but could theoretically die after 1 year and we would observe this death in the data. To balance the fact that some children in the interval may only be observed for 6 months, we will count this as a death within the first year. While the approach is not perfect and will result in slight biases, the impact should be minimal and relatively restricted to the most recent time periods.

\section{Hamiltonian Monte Carlo Algorithm Details}
\label{sec:apd}

We derive the negative log posterior and corresponding gradient for our fertility and mortality models, which are used in a Hamiltonian Monte Carlo (HMC) algorithm \citep{neal:2011} to update the various model parameters.

For the fertility model, define $\bX^{*}_f = \begin{bmatrix}\bX_f, \bZ_f \end{bmatrix}$ where $\bX_f$ is a $n \times p_{X_f}$ matrix ($n$ is the number of observations) and $\bZ_f$ is a $n \times p_{Z_f}$ matrix, and $\btheta^f = \begin{bmatrix}\bbeta^{\top}_f, \bphi^{\top}_f\end{bmatrix}^\top$ where $\bbeta_f$ is a column vector of length $p_{X_f}$ and $\bphi_f$ is a column vector of length $p_{Z_f}$. In the simulation, we do not have $\bZ_f$ or $\bphi_f$. For the hazard model, define $\bX^{*}_h = \begin{bmatrix} \bX_h, \bZ_h\end{bmatrix}$ where $\bX_h$ is a $n \times p_{X_h}$ matrix, and $\bZ_h$ is a $n \times p_{Z_h}$ matrix, and $\btheta^h = \begin{bmatrix}1, \bbeta_h^\top, \bphi_h^\top\end{bmatrix}^\top$ where $\bbeta_h$ is a column vector of length $p_{X_h}$ and $\bphi_h$ is a column vector of length $p_{Z_h}$. Since we will be using a random walk model in time, which does not specify an overall level, we opt to drop the unidentifiable terms from Models (6) and (7) instead of imposing a constraint.

For general $\bX^*$ and $\btheta$, with data model $Y_i\ |\ p_i \sim \text{Binomial}(N_i,k_i p_i)$ where $k_i$ is known (and includes the HIV bias in the mortality model and an adjustment for the final year; see Appendix \ref{sec:apc}) and $\text{logit}(p_i) = \bX_i^{*\top}\btheta$, we have
\begin{align*}
f(\by) & \propto \prod_{i=1}^{n}(k_i p_i)^{y_i}(1-k_ip_i)^{N_i-y_i}\\
\log f(\by) & = \text{const} + \sum_{i=1}^{n}\left\{ y_i\times \text{logit}(k_ip_i) + N_i\times \log(1-k_ip_i) \right\}\\
& = \text{const} +  \sum_{i=1}^{n} \bigg[y_i\times \left(\bX_i^{*\top}\btheta\right) + (N_i-y_i)\times \log\{1+(1-k_i)\exp(\bX_i^{*\top}\btheta)\}\\
 & \qquad - N_i \times \log\left\{1 + \exp(\bX_i^{*\top}\btheta)\right\}\bigg] .
\end{align*}
We assign independent priors,
\begin{equation*}
\begin{aligned}[c]
\bbeta_f & \sim N\big(0,\sigma^2_\beta \mathbf{I}_{p_{X_f}}\big), \\
\bphi_f & \sim N\left(0, (\kappa_f \bK)^{-1}\right),
\end{aligned}
\qquad
\begin{aligned}[c]
\bbeta_h & \sim N\big(0,\sigma^2_\beta \mathbf{I}_{p_{X_h}}\big),\\
\bphi_h & \sim N\left(0, (\kappa_h \bK)^{-1}\right),
\end{aligned}
\end{equation*}
where $\mathbf{I}_p$ is taken to be a $p \times p$ identity matrix, $\sigma_\beta^2=100$, and $\bK$ is the random walk of order 2 (RW2) precision matrix, scaled such that the generalized variance of $\bphi$ is 1, following \cite{sorbye:rue:14}.
We use independent penalized complexity priors for $\kappa_f$ and $\kappa_h$, that is for general $\kappa$,
\begin{align*}
\pi(\kappa) & = \frac{\lambda}{2}\kappa^{-3/2} \exp(-\lambda \kappa^{-1/2}),\\
\lambda & = - \frac{\log(\alpha)}{2}
\end{align*}
where $P(\sigma > u) = \alpha$ with $\sigma = 1/\sqrt{\kappa}$ \citep{simpson:etal:17}. We set $\alpha =0.01$ and $u=0.5$.

Therefore, the negative log posterior (up to a constant) for the hazard model is
\begin{align*}
U_h & = -\by_h^\top \left(\bX_h^*\btheta_h\right) + \bN_h^\top \log\left\{\mathbf{1} + \exp(\bX_h^*\btheta_h)\right\} -(\bN_h-\by_h)^\top\log\{\mathbf{1}+(\mathbf{1}-\bk_h)\circ\exp(\bX_h^*\btheta_h)\} \\
& \qquad + \frac{1}{2\sigma_\beta^2}\bbeta_h^\top \bbeta_h + \frac{\kappa_h}{2}\bphi_h^\top \bK \bphi_h  - \frac{\text{rank}(\bK)}{2} \log(\kappa_h) + \frac{3}{2}\log(\kappa_h) + \lambda \kappa_h^{-1/2}
\end{align*}
where $\by_h$ is a vector containing the number of children that died, $\bN_h$ is a vector containing the number of children at risk, $\bk_h$ is a vector containing the multiplication factors $k_{hi}$, $\text{rank}(\bK)$ is the rank of matrix $\bK$, and $\circ$ denotes element-wise multiplication.
Define $\eta_{\kappa,h} = \log(\kappa_h)$ so that,
\begin{align*}
U_h & = -\by_h^\top \left(\bX_h^*\btheta_h\right) + \bN_h^\top \log\left\{\mathbf{1} + \exp(\bX_h^*\btheta_h)\right\}  -(\bN_h-\by_h)^\top\log\{\mathbf{1}+(\mathbf{1}-\bk_h)\circ\exp(\bX_h^*\btheta_h)\} \\
& \qquad +  \frac{1}{2\sigma_\beta^2}\bbeta_h^\top \bbeta_h + \frac{\exp(\eta_{\kappa,h})}{2}\bphi_h^\top \bK \bphi_h  - \frac{\text{rank}(\bK)}{2} \eta_{\kappa,h}  +\frac{1}{2} \eta_{\kappa,h} + \lambda \exp\left(-\frac{1}{2}\eta_{\kappa,h}\right).
\end{align*}

The gradient is then,
\begin{align*}
\frac{\partial U_h}{\partial \bbeta_h} & = -\by_h^\top \bX_h + \bN_h^\top \{(\bG_h \mathbf{1}^\top_{p_{X_h}}) \circ \bX_h\} - (\bN_h - \by_h)^\top \{(\bG_{h,k}\mathbf{1}_{p_{X_h}}^\top)\circ \bX_h\} + \frac{1}{\sigma_\beta^2}\bbeta_h^\top\\
\frac{\partial U_h}{\partial \bphi_h} & = -\by_h^\top \bZ_h + \bN_h^\top \{(\bG_h \mathbf{1}_{p_{Z_h}}^\top) \circ \bZ_h\} - (\bN_h - \by_h)^\top \{(\bG_{h,k}\mathbf{1}_{p_{Z_h}}^\top)\circ \bZ_h\} + \exp(\eta_{\kappa,h}) \bphi_h^\top \bK\\
\frac{\partial U_h}{\partial \eta_{\kappa,h}} & = \frac{\exp(\eta_{\kappa,h})}{2}\bphi_h^\top \bK \bphi_h -\frac{\text{rank}(\bK)}{2} + \frac{1}{2} - \frac{1}{2}\lambda \exp\left(-\frac{1}{2}\eta_{\kappa,h}\right)
\end{align*}
where $\bG_h = \text{expit}(\bX^*_h\btheta_h)$ (a column vector of length $n$) and $\bG_{h,k} = \frac{(\mathbf{1}-\bk_h)\ \circ\ \exp(\bX^*_h\btheta_h)}{\mathbf{1} + (\mathbf{1}-\bk_h)\ \circ\ \exp(\bX_h^\star \btheta_h)}$ (also a column vector of length $n$).

The negative log posterior and gradient is similar for the fertility model. In the simulation the vector $\bk$ is set to 1.

\section{Aggregation Algorithm}
\label{sec:ape}

Define $m$ to be woman's age, $s$ to be strata (rural, urban), $d$ to be district, $t$ to be the year, and $p$ to be the 5-year time period. Define $\text{FP}(m; t; d; s)$ to be the total number of women of age $m$ at time $t$ in district $d$ and strata $s$. Denote the number of births these women experience (Children Ever Born) as $\text{CEB}(m;t;d;s)$. Let $j=1,\dots,J$ index the samples. To obtain samples of $q(5)$ on the 5-year time period and district level, we employ the following steps:
\begin{enumerate}
\item Simulate $\text{CEB}(m;t;d;s)^{(j)} \sim \text{Binomial}(\text{FP}(m;t;d;s), f(m;p;d;s)^{(j)})$
\item Transform:
\begin{align*}
		_5 q_0(p;d)^{(j)} = \sum_{t\in p} \sum_{s} {_5q_0(t;d;s)^{(j)}}\left\{ \frac{\sum_m \text{CEB}(m;t;d;s)^{(j)}}{\sum_{t \in p}\sum_s\sum_m \text{CEB}(m;t;d;s)^{(j)}}\right\}.
		\end{align*}
\end{enumerate}

\section{Additional Results for Malawi Application}
\label{sec:apf}

\begin{figure}[b]
\centerline{\includegraphics[width=6.5in]{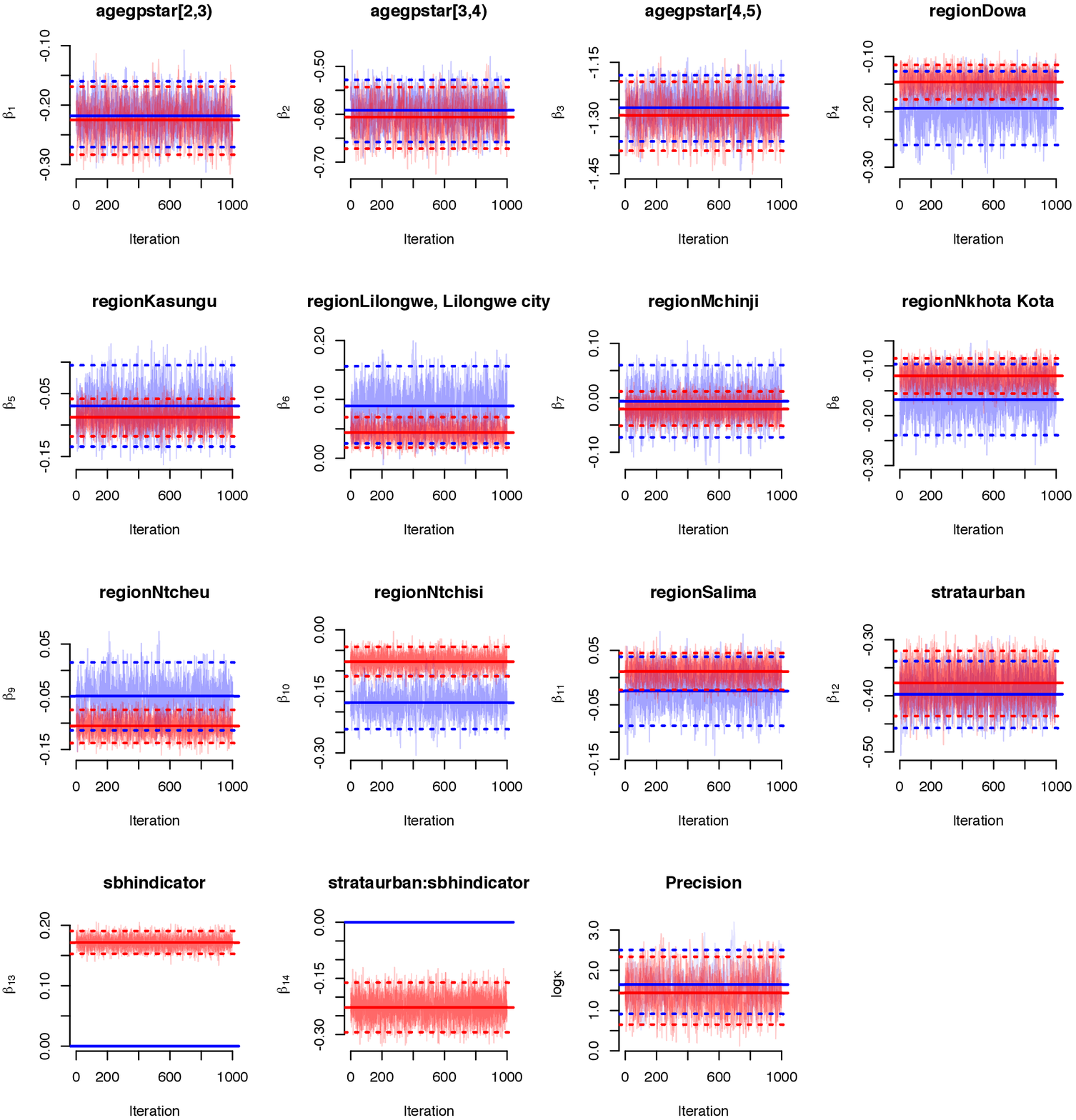}}
\caption{Trace plots for parameters in the mortality model, $\beta_{c[a]}$, $\beta_{district}$, $\beta_{strata}$, $\beta_{SBH,strata}$, and precision parameter $\kappa$ for the RW2. Red are results from the FBH + SBH model. Blue are results from the FBH only model. Solid lines: posterior medians. Dashed lines: 95\% CI.}
\vspace*{-3pt}
\label{Fig:tracehaz}
\end{figure}

Trace plots for the fixed effects and precision of the RW2 models suggest convergence (Figure \ref{Fig:tracehaz}). 
Posteriors for the fertility parameters look similar. Posteriors for the hazards look similar  for the age of the child, $\beta_{c[m]}$. However, we do see some differences in posteriors for the district-level effects. In terms of the urban effect, the posterior medians are -0.397 (95\% CI: -0.458, -0.338)  for FBH data only and -0.377 (95\% CI: -0.458, -0.338) for FBH + SBH. In terms of the bias terms, the posterior median for the indicator for SBH data is 0.171 (95\% CI: 0.153, 0.191) and for the indicator for SBH and urban data is -0.228 (95\% CI: -0.294, -0.162). The interpretation is that the SBH data result in an increase in the yearly hazard odds of 18.6\% in rural areas and a decrease in the yearly hazard odds of 5.5\%.



\begin{figure}[b]
\centerline{\includegraphics[width=6in]{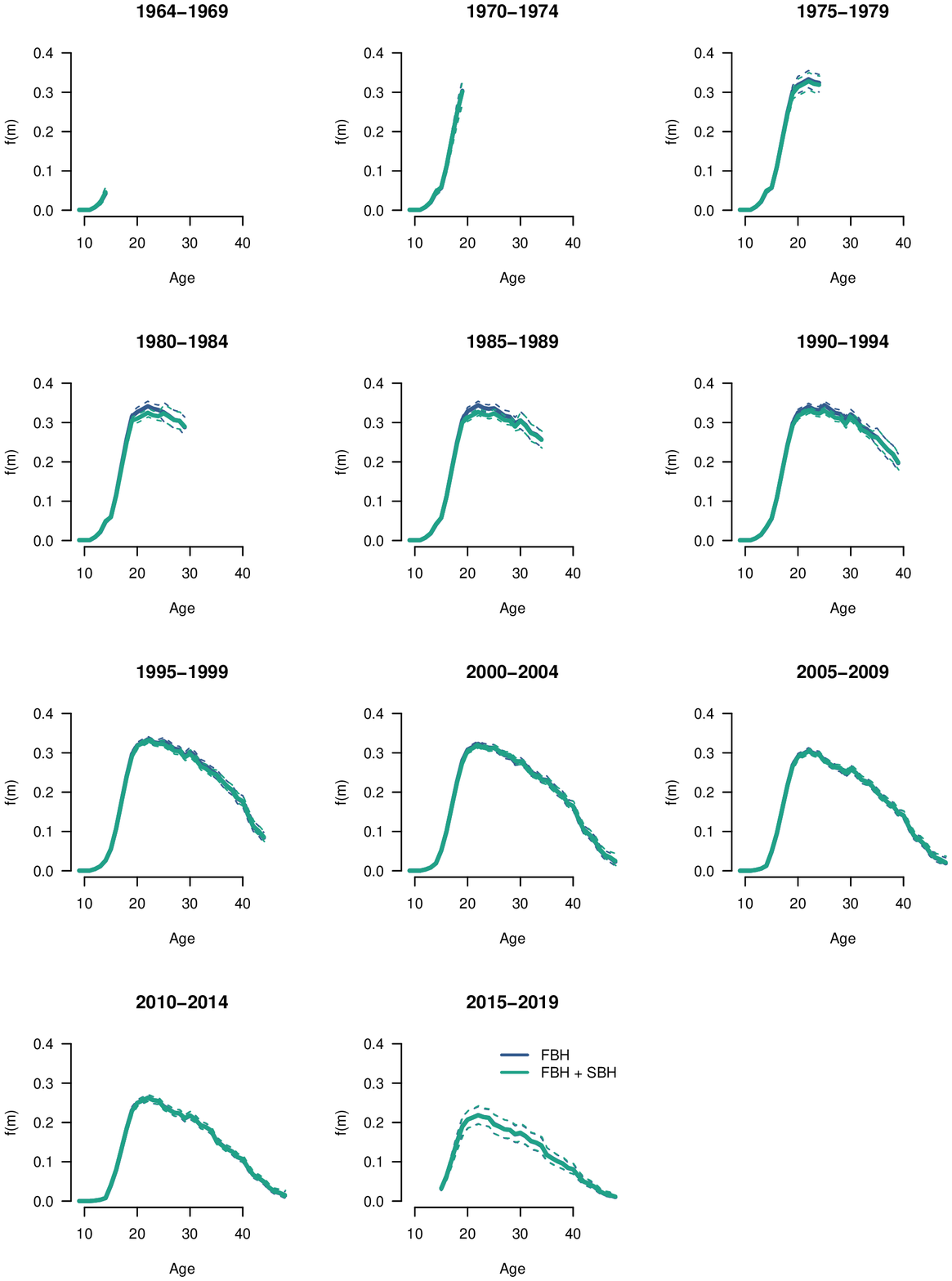}}
\caption{Fertility probabilities in rural areas by age of woman across 5-year time periods. Solid lines: posterior medians. Dashed lines: 95\% credible intervals. Ages where there are no observations are excluded.}
\vspace*{-3pt}
\label{Fig:fertr}
\end{figure}

\begin{figure}[b]
\centerline{\includegraphics[width=6in]{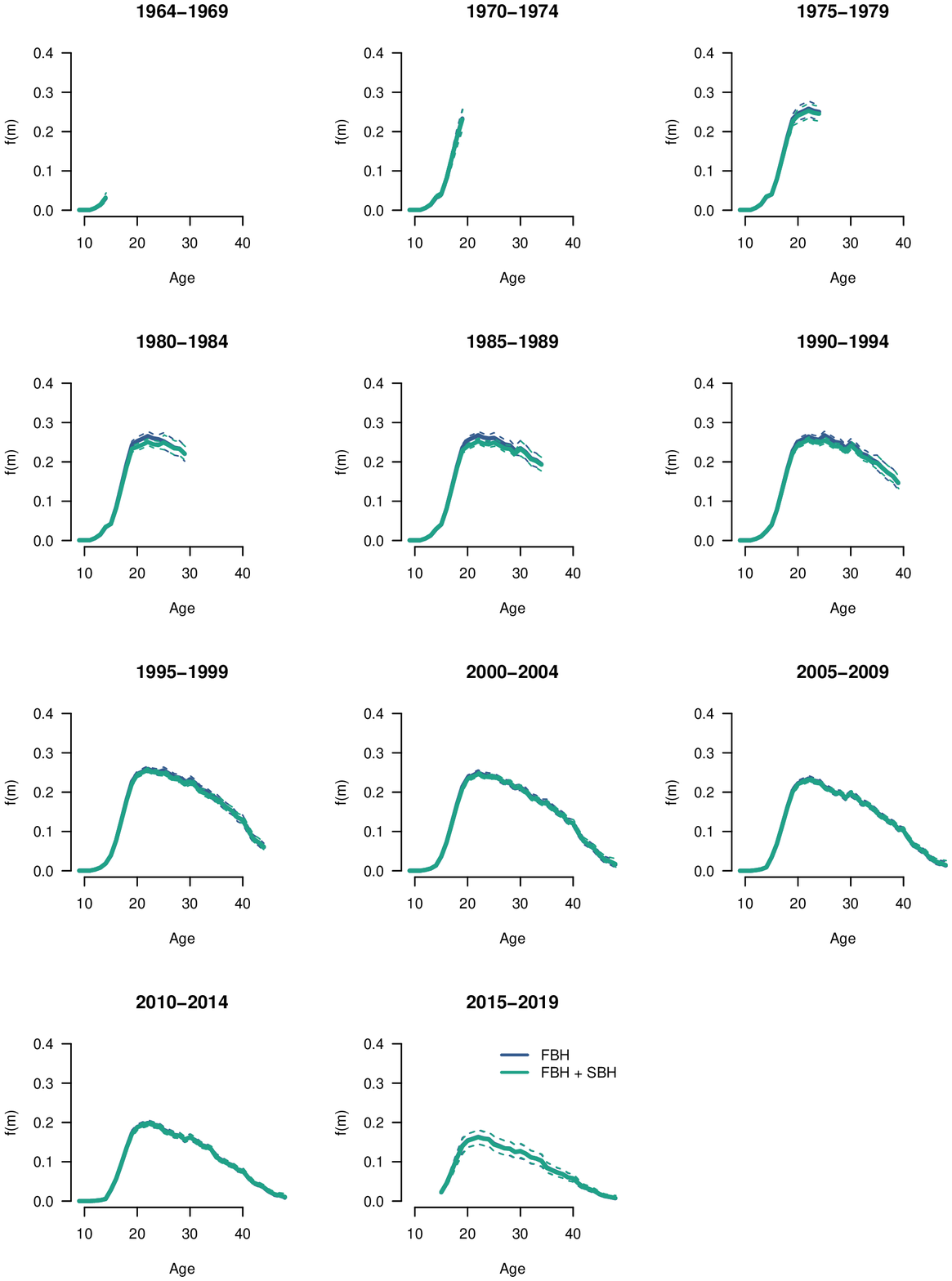}}
\caption{Fertility probabilities in urban areas by age of woman across 5-year time periods. Solid lines: posterior medians. Dashed lines: 95\% credible intervals. Ages where there are no observations are excluded.}
\vspace*{-3pt}
\label{Fig:fertu}
\end{figure}

\begin{figure}[b]
\centerline{\includegraphics[width=6.5in]{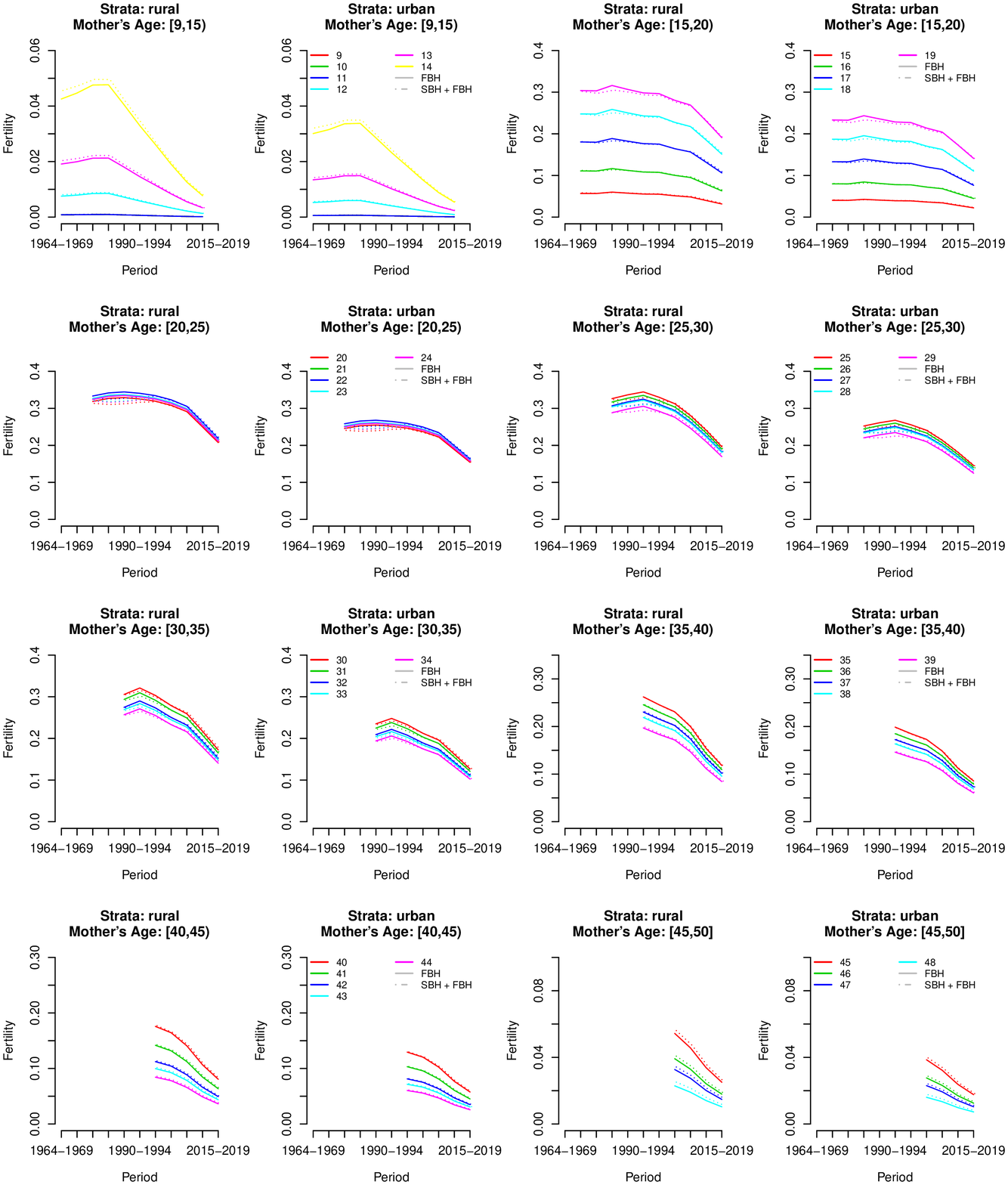}}
\caption{Fertility probabilities (posterior medians) by 5-year time periods across woman's age. Ages where there are no observations are excluded.}
\vspace*{-3pt}
\label{Fig:fert}
\end{figure}

Fertility probabilities are similar when SBH data are added (Figures \ref{Fig:fertr}--\ref{Fig:fert}). Fertility tends to be higher in rural areas, to be highest among women in their twenties, and to decrease over time for all age groups.

\begin{figure}[b]
  \centerline{\includegraphics[width=6.5in]{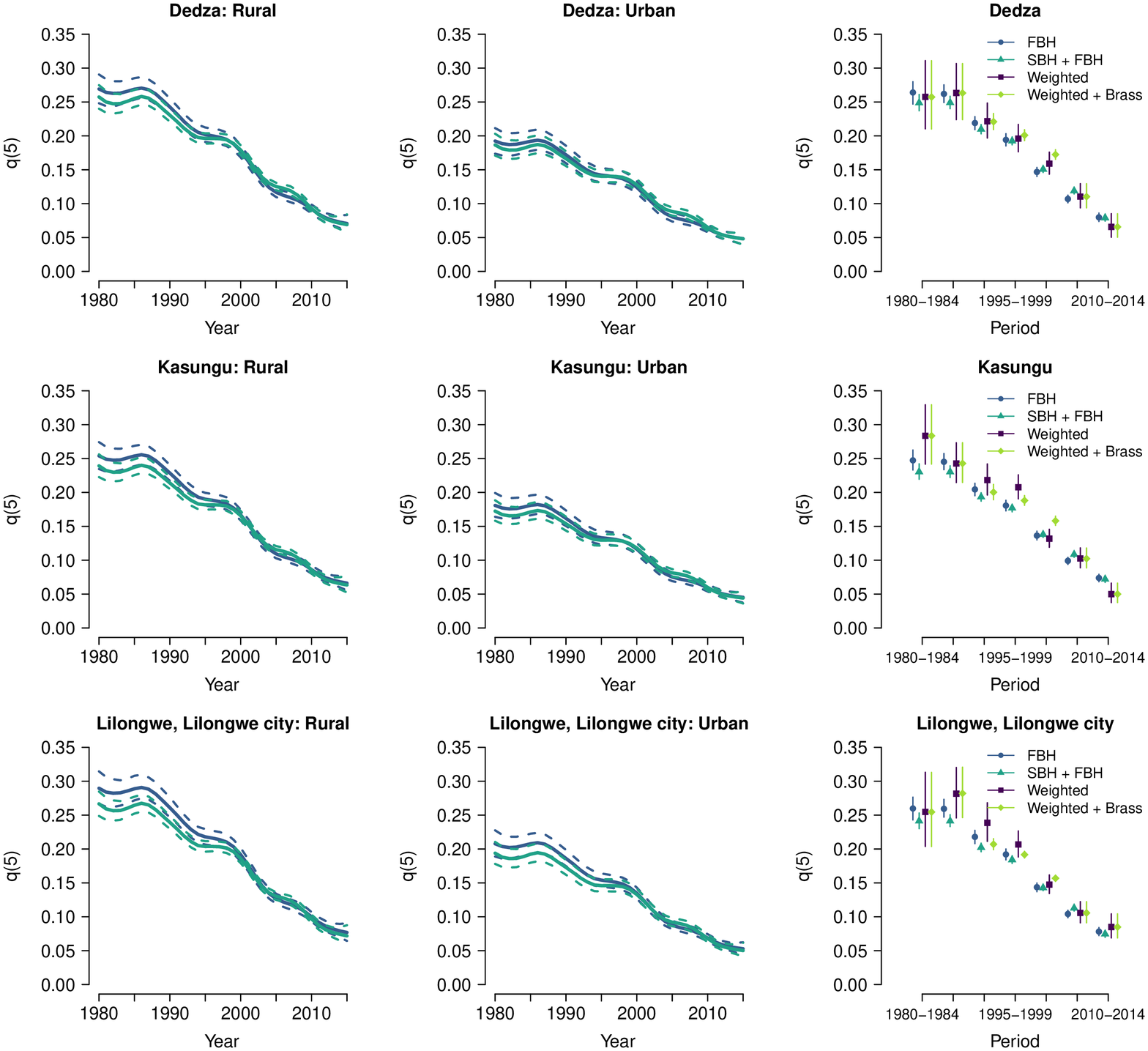}}
\caption{Left and Middle Panels: Posterior medians (solid lines) and 95\% credible intervals (dashed lines) for U5MR ($q(5)$) in 3 districts. Right Panel: Point estimates and 95\% uncertainty interval for the 4 different methods.}
\vspace*{-3pt}
\label{Fig:q5a}
\end{figure}

\begin{figure}[b]
  \centerline{\includegraphics[width=6.5in]{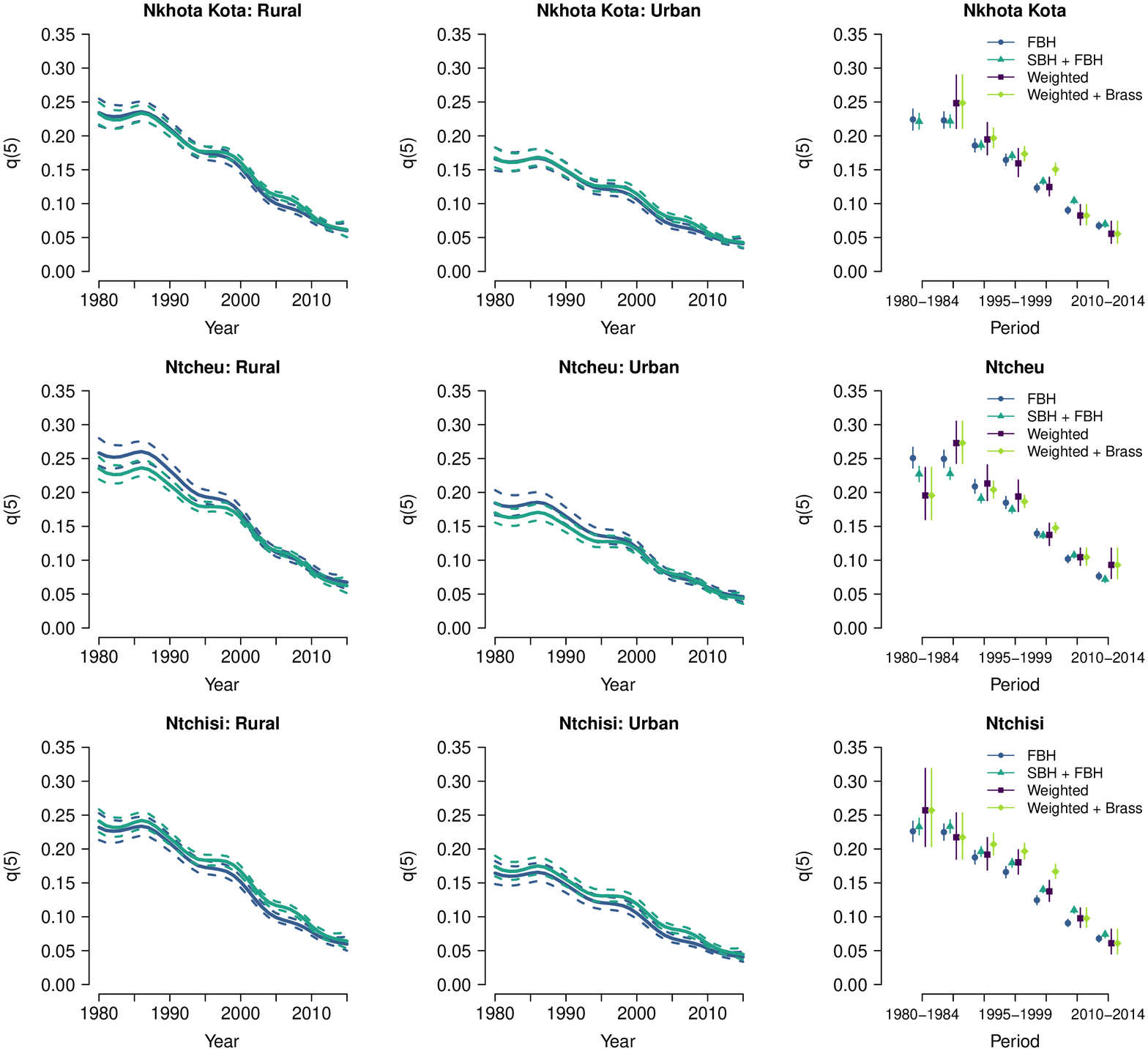}}
\caption{Left and Middle Panels: Posterior medians (solid lines) and 95\% credible intervals (dashed lines) for U5MR ($q(5)$) in 3 districts. Right Panel: Point estimates and 95\% uncertainty interval for the 4 different methods. For Nkhota Kota, the results from the weighted methods fall outside the plotting region.}
\vspace*{-3pt}
\label{Fig:q5b}
\end{figure}

Results for U5MR comparing the 4 methods in the remaining 6 districts are presented in Figures \ref{Fig:q5a} and \ref{Fig:q5b}.

\begin{figure}[b]
  \centerline{\includegraphics[width=6.5in]{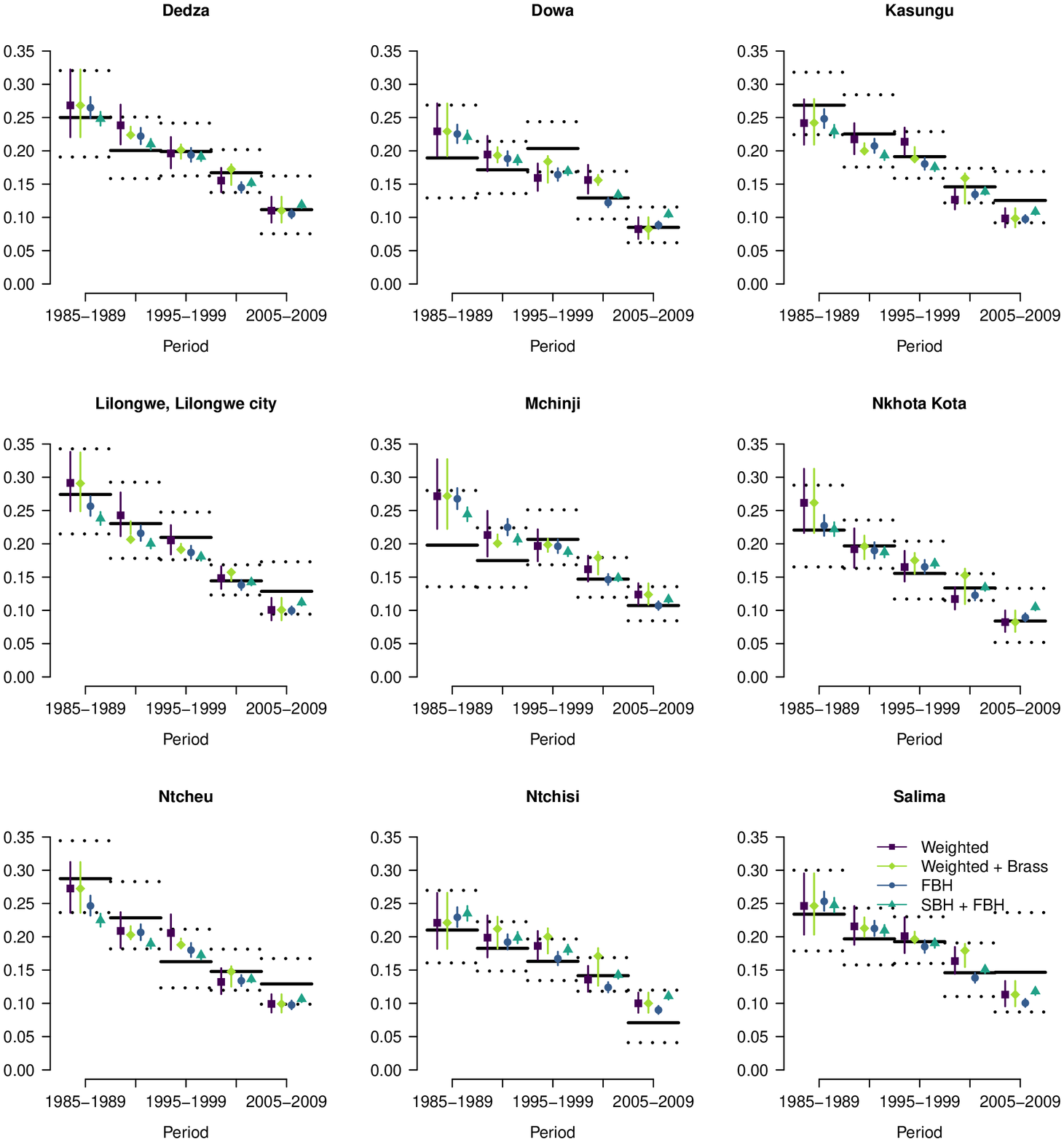}}
\caption{Solid black lines are the direct estimates obtained from the holdout data and the dashed black lines are the corresponding 95\% confidence intervals. Points are $\text{expit}\left(E [Y_{dp}^{(M)}]\right)$ and the vertical lines are the corresponding 95\% uncertainty intervals constructed using $\text{Var}(Y_{dp}^{(M)})$ and the quantiles of a normal distribution and then transformed to the probability scale.}
\vspace*{-3pt}
\label{Fig:mse}
\end{figure}

To assess accuracy of the various models we fit the models to the training data and compared the results to the direct estimates obtained from the holdout data. Define $Y_{dp}^{(M)}$ to be the logit of U5MR in district $d$ time period $p$. Figure \ref{Fig:mse} provides a sense of the uncertainty in the numbers used as the ``truth'' and displays the estimate and 95\% uncertainty interval of $Y_{dp}^{(M)}$ for each model $M$ on the probability scale, where we approximate the distribution of $Y_{dp}^{(M)}$ with a normal distribution.

\begin{figure}[b]
  \centerline{\includegraphics[width=6.5in]{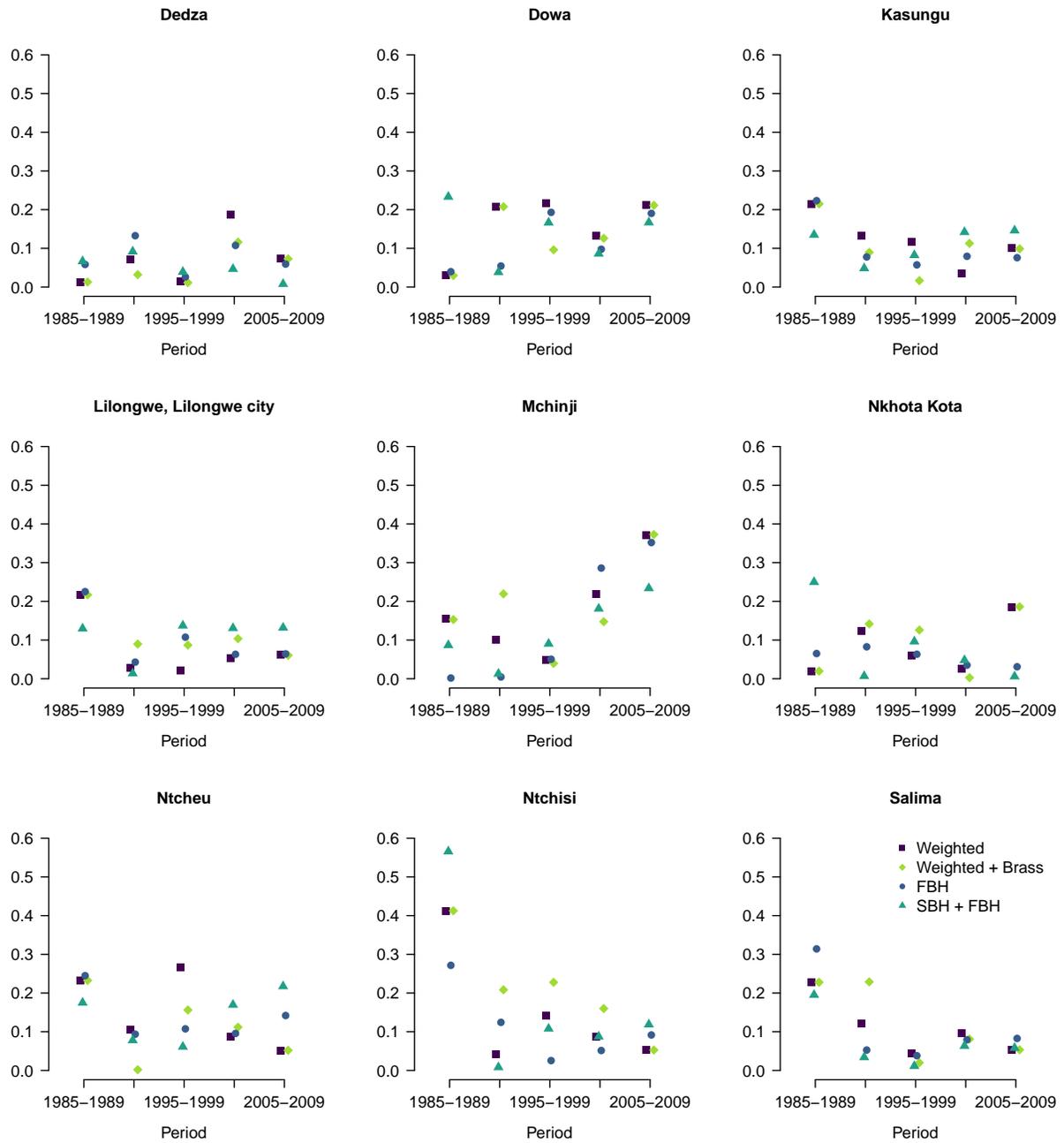}}
\caption{Percent absolute relative error by district and period.}
\vspace*{-3pt}
\label{Fig:pare}
\end{figure}

We also calculate the percentage relative error (PARE; Table \ref{Tab:pare}),
\begin{align*}
\text{PARE}(p)^{(M)} = \frac{1}{9} \sum_{d=1}^9 w_{dp} \frac{|\hat{q}_{dp}(5)^{(M)} - q_{dp}(5)|}{q_{dp}(5)}
\end{align*}
where $w_{dp} = \hat{V}_d(p)^{-1}/\sum_{d=1}^9 \hat{V}_d(p)^{-1}$ and $\hat{V}_d(p)^{-1}$ is the variance of the direct estimates. These (unweighted) values are shown for each district $d$ and period $p$ in Figure \ref{Fig:pare}.

\begin{table}[b]
\caption{Percent Absolute Relative Error of $q(5)$ estimates ($\times 100\%$) by Model.}
\label{Tab:pare}
\begin{center}
\begin{tabular}{lrrrr}
\hline
Period & Weighted Estimates & Weighted Estimates  & Smoothed Model: & Smoothed Model: \\
& & + Brass & FBH & FBH + SBH \\
\hline
1985--1989 & 10.5 & 10.4 & 10.4 & 12.4 \\  
1990--1994 & 9.82 & 10.2 &  9.21 &  9.91 \\
1995--1999 & 9.69 & 8.18 & 7.42  & 9.03 \\  
2000--2004 & 9.41 & 12.7 &  7.62 & 3.24 \\  
2005--2009 & 16.2 & 16.2 & 13.8 & 16.7 \\
Average &  10.5 & 11.1 & 8.96 & 8.76 \\
\hline
\end{tabular}
\end{center}
\end{table}

\end{document}